\documentclass[journal]{IEEEtran}
\usepackage{xcolor,soul,framed} 

\colorlet{shadecolor}{yellow}
\usepackage[pdftex]{graphicx}
\graphicspath{{../pdf/}{../jpeg/}}
\DeclareGraphicsExtensions{.pdf,.jpeg,.png}

\usepackage[cmex10]{amsmath}
\usepackage{array}
\usepackage{mdwmath}
\usepackage{mdwtab}
\usepackage{eqparbox}
\usepackage{url}
\usepackage{svg} 

\usepackage{alphabeta}
\usepackage[acronym,nomain]{glossaries}
\usepackage{textcomp,gensymb}
\usepackage{floatrow}
\usepackage[label font=bf,labelformat=simple]{subfig}
\usepackage{esvect}
\usepackage{booktabs}
\usepackage{csquotes}
\usepackage{multicol} 
\usepackage[font=footnotesize]{caption}
\usepackage[url=false, style=ieee,doi=false,isbn=false,]{biblatex}
\AtBeginBibliography{\footnotesize}
\usepackage{makecell} 
\begin{document}

    \title{Ultrafast 3-D Super Resolution Ultrasound using Row-Column Array specific Coherence-based Beamforming and Rolling Acoustic Sub-aperture Processing: In Vitro, In Vivo and Clinical Study}
  \author{Joseph Hansen-Shearer, Jipeng Yan, Marcelo Lerendegui, Biao Huang, Matthieu Toulemonde, Kai Riemer, Qingyuan Tan, Johanna Tonko, Peter D. Weinberg, Chris Dunsby, Meng-Xing Tang
\vspace{-0.5cm}
  \thanks{We would like to acknowledge the funding from the EPSRC CDT in Smart Medical Imaging, the EPSRC project grant (EP/T008970/1), the National Institute for Health Research i4i under Grant NIHR200972, the Chan Zuckerberg Foundation under Grant No. 2020-225443, and the Department of Bioengineering at Imperial College London. Corresponding Email: mengxing.tang@imperial.ac.uk}}

\maketitle
\newacronym{rca}{RCA}{Row-Column Array}
\newacronym{das}{DAS}{Delay and Sum}
\newacronym{rc-fmas}{RC-FMAS}{Row Column specific Frame Multiply and Sum}
\newacronym{psf}{PSF}{Point Spread Function}
\newacronym{fwhm}{FWHM}{Full Width Half Maximum}
\newacronym{smer}{SMER}{Side-to-Main Lobe Energy Ratio}
\newacronym{psmr}{PSMR}{Peak Side-to-Main Lobe Ratio}
\newacronym{cr}{CR}{Contrast Ratio}
\newacronym{gcnr}{gCNR}{generalised Contrast to Noise Ratio}
\newacronym{snr}{SNR}{Signal to Noise Ratio}
\newacronym{rf}{RF}{Radio Frequency}
\newacronym{ulm}{ULM}{Ultrasound Localisation Microscopy}
\newacronym{opwc}{OPWC}{Orthogonal Plane Wave Compounding}
\newacronym{asap}{ASAP}{Acoustic Sub-Aperture Processing}
\newacronym{svd}{SVD}{Singular Value Decomposition}
\newacronym{mip}{MIP}{Maximum Intensity Projection}
\newacronym{fsc}{FSC}{Fourier Shell Correlation}
\begin{abstract}

The row-column addressed array is an emerging probe for ultrafast 3-D ultrasound imaging. It achieves this with far fewer independent electronic channels and a wider field of view than traditional 2-D matrix arrays, of the same channel count, making it a good candidate for clinical translation. However, the image quality of row-column arrays is generally poor, particularly when investigating tissue. Ultrasound localisation microscopy allows for the production of super-resolution images even when the initial image resolution is not high. Unfortunately, the row-column probe can suffer from imaging artefacts that can degrade the quality of super-resolution images as `secondary' lobes from bright microbubbles can be mistaken as microbubble events, particularly when operated using plane wave imaging. These false events move through the image in a physiologically realistic way so can be challenging to remove via tracking, leading to the production of 'false vessels'. Here, a new type of rolling window image reconstruction procedure was developed, which integrated a row-column array-specific coherence-based beamforming technique with acoustic sub-aperture processing for the purposes of reducing `secondary' lobe artefacts, noise and increasing the effective frame rate. Using an {\it{in vitro}} cross tube, it was found that the procedure reduced the percentage of `false' locations from $\sim$26\% to $\sim$15\% compared to traditional orthogonal plane wave compounding. Additionally, it was found that the noise could be reduced by $\sim$7 dB and that the effective frame rate could be increased to over 4000 fps. Subsequently, {\it{in vivo}} ultrasound localisation microscopy was used to produce images non-invasively of a rabbit kidney and a human thyroid.


\end{abstract}

\begin{IEEEkeywords}
3-D Ultrasound, high frame rate / ultrafast imaging, real-time,  Row-Column, Beamforming, Coherence, Frame multiply and Sum
\end{IEEEkeywords}

\section{Introduction}

Super-resolution ultrasound through localising and tracking microbubbles, also known as \gls{ulm}, is a technique capable of breaking the diffraction limit of an ultrasound system ~\cite{christensen-jeffries_super-resolution_2020, christensen-jeffries_vivo_2015, errico_ultrafast_2015}. In more recent years, super-resolution has also been achieved in 3-D~\cite{demeulenaere_coronary_2022,heiles_ultrafast_2019,foroozan_microbubble_2018,christensen-jeffries_3-d_2017}. Achieving good quality super-resolution in 3-D and {\it{in vivo}} can be challenging due to the small field of view, typically higher acquisition times, lower initial image quality and large quantities of received data compared to 2-D. Usually with research systems, the acquisition needs to pause while a large amount of data is saved to the computer or the frame rate needs to be reduced to enable continuous saving. In a clinical setting, this is particularly problematic as the majority of ultrasound contrast agents used get destroyed by the body within a few minutes. Multiple injections for longer acquisition are possible, but holding the probes steady for many minutes without the patient or probe moving can be challenging. By reducing the channel count, and thus the quantity of data, longer acquisitions with less disruption become possible.

\begin{figure} [htb]
    \centering
        \includesvg[inkscapelatex=false,width=1\columnwidth]{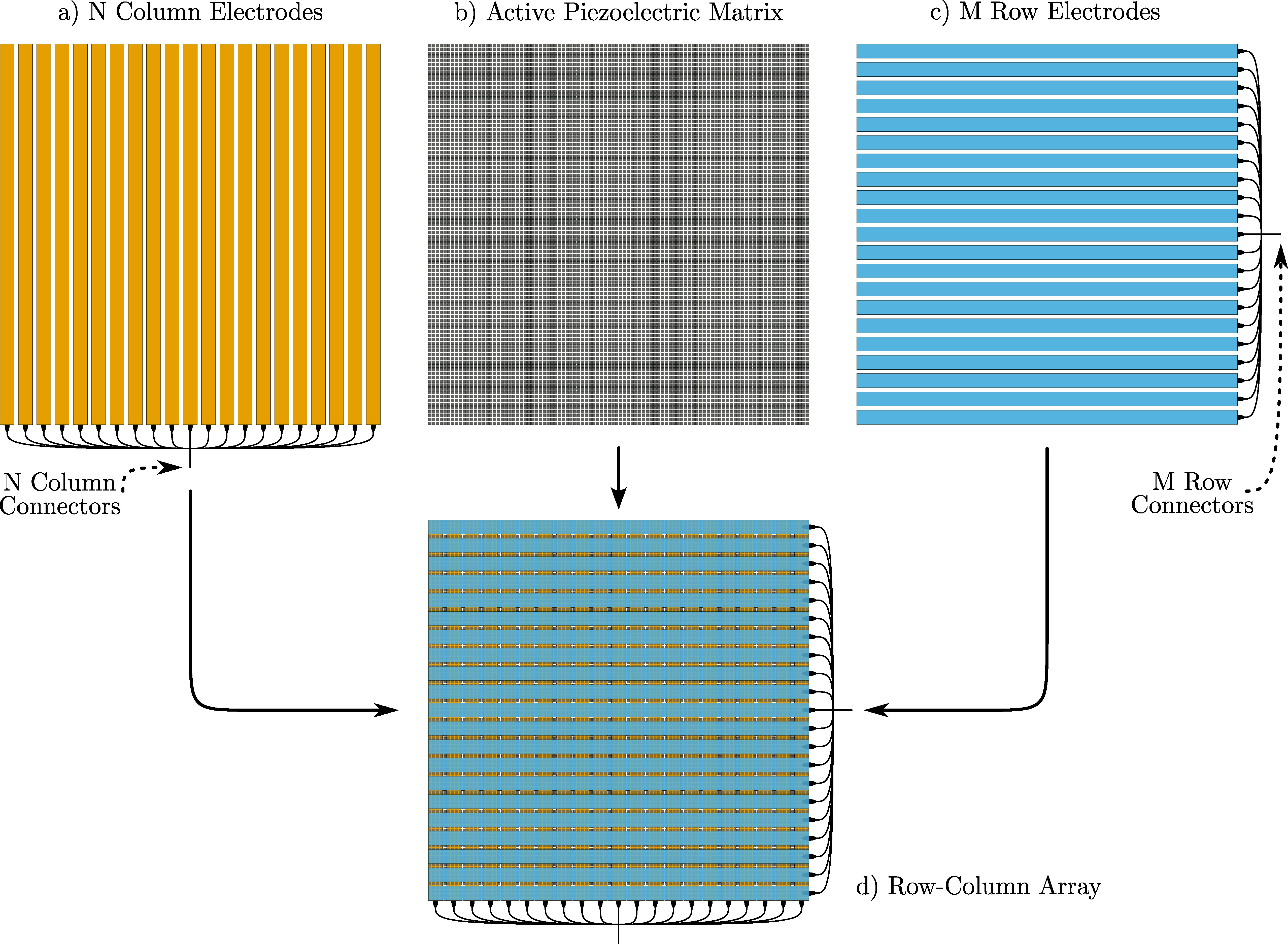}
        \caption{Diagram showing the basic configuration of the row-column array transducer used in this work. Here, orange represents the active area of row transmissions (a) and turquoise represents column transmissions (c). The two arrays of electrodes stimulate an active matrix array of piezoelectric elements (b) placed between them. The total number of elements has been reduced here for readability.}
        \label{fig:RCA_Diagram}
        
\end{figure}

The \gls{rca} ultrasound probe is a newly emerging technology able to produce large field of view 3-D ultrasound images with a reduced channel count compared to fully addressed matrix arrays~\cite{jensen_three-dimensional_2019}. In this work, it consisted of two overlapping series of driving elements aligned orthogonally to each other (rows and columns). In between these two layers, a bed of piezoelectric elements is tiled. Instead of the traditional point-like elements, all the elements are elongated to form long thin elements in such a way that their footprint covers the entire probe surface, see Figure~\ref{fig:RCA_Diagram}~\cite{ferin_ultrasound_2019}. This configuration allows for a dramatic reduction of the number of electrical channels from $R\times C$ to $R + C$ at the cost of limiting beam steering and reception to two orthogonal directions. The \gls{rca} has a number of artefacts that can make imaging {\it{in vivo}} challenging, particularly when operating in the ultrafast regime, which is often needed for super-resolution imaging. Great progress has been made with this type of probe \textit{in silico} and \textit{in vitro}~\cite{seo_5a-5_2006,savoia_p2b-4_2007, pappalardo_bidimensional_2008, demore_real-time_2009,sampaleanu_top-orthogonal--bottom-electrode_2014,rasmussen_3-d_2015,christiansen_3-d_2015,li_preliminary_2015,holbek_3-d_2016,bouzari_curvilinear_2017,chen_column-row-parallel_2018,ceroici_fast_2019,jensen_three-dimensional_2019}. \textit{In vivo}, a humeral artery and a carotid artery were imaged with power Doppler~\cite{flesch_4d_2017,sauvage_large_2018}, a 3-D functional brain image on an exposed rat brain~\cite{sauvage_4d_2020} and a super-resolution image of a surgically exposed rat kidney~\cite{jensen_anatomic_2022} was produced. 

In super-resolution imaging, contrast agents (microbubbles) are localised and tracked as they flow through the body. To perform this, particularly in vessels with faster flow, a high frame rate is required. This is even more important if the concentration of microbubbles is high, as in order to build up an image of the vasculature, assumptions about the location of microbubbles between frames are required. If there is large inter-frame displacement (low frame rate), then these assumptions will no longer be valid and thus the accuracy of the reconstructions will decrease.

An ultrafast imaging technique, \gls{opwc}, was developed for \gls{rca} by M. Flesch et al. ~\cite{flesch_4d_2017}. Although this technique can produce ultrafast ultrasound images, it suffers from some strong imaging artefacts. In \gls{opwc}, a series of volumes are produced with transmissions steered either with the row or the column elements and then reception is performed with the orthogonal series of elements. The \gls{psf} produced from each transmission-reception event is elongated in the direction of row and column elements, respectively. This is due to the lack of two-way elevation and azimuth focusing. The volumes are then coherently summed, which produces the full 3-D information required to form an image. This process produces a final \gls{psf} that has a {\it{`cross-like'}} structure. By increasing the number of steered transmission angles, the {\it{`cross-like'}} nature of the \gls{psf} can be reduced but not fully eliminated, resulting in outlying `secondary' lobes on all four sides of the \gls{psf}. The fact that this problem is particularly evident with fewer transmission angles is problematic for super-resolution imaging, where a smaller number of transmission angles is desirable. 

In this work, we demonstrate non-invasive 3-D super-resolution imaging using a \gls{rca} probe and ultrafast imaging. {\it{In vivo}}, ultrasound localisation microscopy was applied to a rabbit kidney and a human
thyroid, integrating a \gls{rc-fmas} technique~\cite{hansenshearer_rc_fmas} with~\gls{asap}~\cite{stanziola_asap_2018} and rolling processing, explained below. These organs were chosen for demonstration purposes only. 
\gls{rc-fmas} is a coherence-based ultrafast imaging technique that aims to improve the quality of the \gls{psf} produced when imaging the contrast agents used in the \gls{ulm} procedure. \gls{asap} is implemented to reduce noise, and rolling processing is used to increase the effective frame rate and hence improve \gls{ulm} tracking. By combining all three, noise can be reduced without sacrificing temporal resolution, which would typically be the case when performing \gls{asap}. Prior to the {\it{in vivo}} studies, a cross tube was imaged {\it{in vitro}} to demonstrate the problems associated with high `secondary' lobes and how they can be rectified via \gls{rc-fmas}.

\section{Methods}
To produce the super-resolution image, the following acquisition and processing pipeline was used. The region of interest was imaged using a series of row and column-based transmissions followed by orthogonally-based receptions. The received signal was then IQ demodulated and downsampled. Clutter was removed via \gls{svd}~\cite{demene_spatiotemporal_2015} on the complex channel data, with each transmission angle being processed separately. The volumes were then reconstructed via rolling reconstruction using \gls{opwc} and \gls{rc-fmas} with and without \gls{asap}. These four techniques were then compared. In the rabbit experiments, rolling reconstruction and \gls{asap} were compared to traditional \gls{opwc} without rolling reconstruction or \gls{asap}. In the {\it{in vivo}} cases, motion correction was applied using the Matlab (The MathWorks, Inc., Natick, MA, USA) function {\it{imregdemons}}. In the cross-tube experiment, this step was skipped. Localisation was then performed via auto-thresholding and tracking was performed via Kalman filtering~\cite{yan_super-resolution_2022}. All experiments were performed using a Verasonics Vantage 256 platform (Verasonics Inc., Redmond, WA). The probe used was a 128 + 128 element \gls{rca} and with an active footprint of 26.4 $\times$ 26.4 mm \cite{ferin_ultrasound_2019} supplied by Vermon S.A. (Tours, France). A 3 MHz pulse was used for all transmissions, the sampling frequency was 12 MHz and the pulse repetition frequency for all, imaging depth was selected as maximum depth shown in each figure and the pulse repetition frequency was set as high as achievable given these depths.
The acquisition parameters can be found in Table~\ref{tab:chapter:proccessing_paramters}. During all experiments the angular pitch was set at 1.1$\degree$, to ensure grating lobes would be outside the field of view~\cite{montaldo_coherent_2009,denarie_coherent_2013,sauvage_large_2018}.

\subsection{Clutter filtering}
\gls{svd} was used to remove the clutter signal from the cross tube and {\it{in vivo}} tissue, rather than pulse encoding which is sometimes used for super-resolution imaging. This was done for two reasons. First, \gls{rca} requires twice as many transmissions as a linear probe, due to its requirement for steering in two directions; thus further reduction in the frame rate is not desirable. Second, the linearity of the imaging system and probe used in this project was not high, so techniques such as amplitude modulation~\cite{brock-fisher_means_1996} and pulse inversion~\cite{hwang_two_1999} did not provide sufficient tissue reduction. \gls{svd} is a memory-intensive operation that utilises many frames at once. After beamforming, the size of the data increases significantly, when using a \gls{rca}. To reduce the computational cost, it was performed prior to beamforming, on each angle separately, using the IQ-demodulated \gls{rca} data.

\subsection{Beamforming}
\subsubsection{Orthogonal Plane Wave Compounding}
\begin{figure} [htb]
    \centering
        \includegraphics[width=0.95\linewidth]{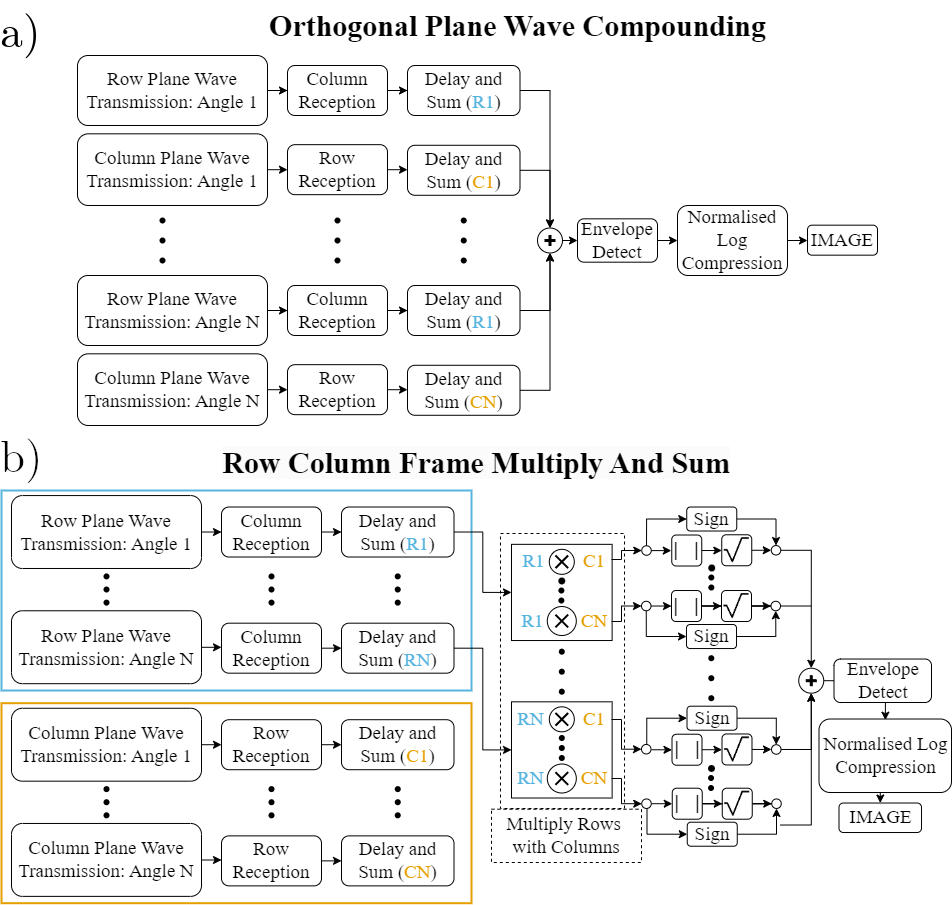}
        \caption{a) Diagram of the \gls{opwc} image reconstruction procedure. b) Diagram of the \gls{rc-fmas} image reconstruction procedure. }
        \label{fig:OPWC_RC-FMAS_Diagram}
\end{figure}

\gls{opwc} was used as the baseline. A modified \gls{rca} specific~\gls{das} algorithm~\cite{rasmussen_3-d_2015} was used to generate a series of volumes ($V_{DAS}$). After this, coherent compounding can be used to generate a final volume. The \gls{opwc} reconstruction pipeline is shown in Figure~\ref{fig:OPWC_RC-FMAS_Diagram} a.

\subsubsection{Row-Column specific Frame Multiply and Sum}
\gls{rc-fmas} is a coherence-based reconstruction technique that has been shown to reduce the level of the `secondary' lobes generated in \gls{opwc} by over 16 dB~\cite{hansenshearer_rc_fmas}. It acquires data and beamforms a series of volumes in the same way as \gls{opwc}. However, instead of simply coherently compounding the volumes, the volumes are combined into a series of pairs. Each row transmission is paired with each column transmission with the total number of pairings thus being

\begin{equation}
N_P' = N_{R_{tx}} \cdot N_{C_{tx}},
\label{eq:fmas_rc_np}
\end{equation}

where $N_{R_{tx}}$ and $N_{C_{tx}}$ are the number of rows transmissions and columns transmissions respectively. A new volume ($V_{RCFMAS}$) is then calculated by multiplying each of these pairs together, extracting the phase, taking the square root of each pair to maintain intensity linearity (homogeneity of degree one), and then reapplying the phase. After this, the pairwise combinations are coherently compounded. The final volume is thus:

\begin{equation}
V_{RCFMAS}(\vv{\bf{r}}) = \sum_{i= 1}^{N_{R_{Tx}}} \sum_{j=1}^{N_{C_{Tx}}} sign(V_{ij}(\vv{\bf{r}})) \cdot \sqrt{\Big\vert V_{ij}(\vv{\bf{r}})\Big\vert} ,
\end{equation}
where $V_{ij}$ represents each row-column volume pairing. The pipeline for generating images with \gls{rc-fmas} is shown in Figure~\ref{fig:OPWC_RC-FMAS_Diagram} b.

\begin{figure*} [ht]
    \centering
        \includegraphics[width=1\linewidth]{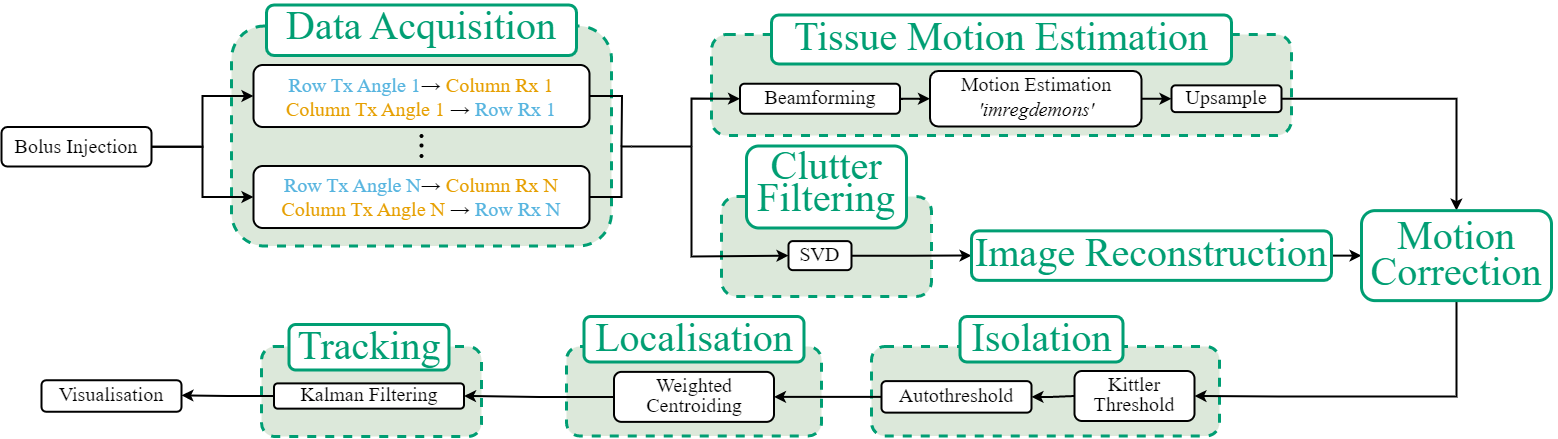}
        \caption{Shows the processing pipeline used to produce super-resolved volumes. For the cross-tube experiment, the bolus injections were replaced with an infusion pump and no motion compensation was needed. Image reconstruction here is a stand-in for the four techniques: \gls{opwc} or \gls{rc-fmas} both with and without \gls{asap} with rolling compounding. The beamforming step in the tissue motion estimation was \gls{rc-fmas} without rolling reconstruction.}
        \label{fig:processing_pipeline}
        
\end{figure*}
\subsection{Rolling Reconstruction}

\begin{figure} [htb]
    \centering
        \includegraphics[width=\linewidth]{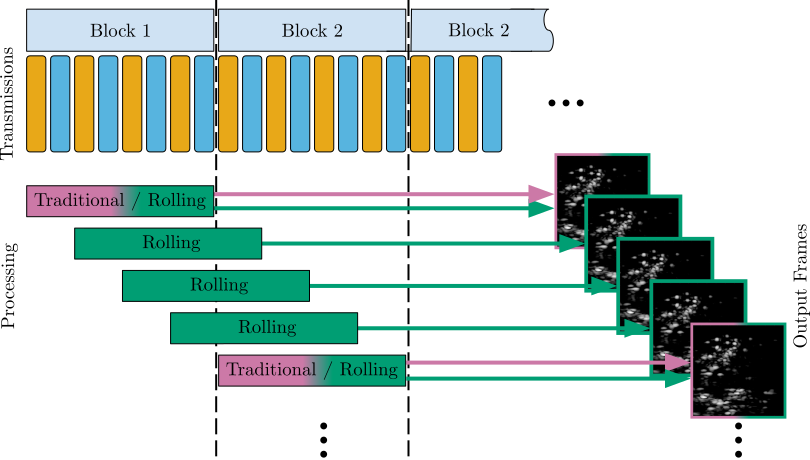}
        \caption{Schematic showing the rolling processing. The orange and blue boxes represent the row and column transmission-reception events. For simplicity, only 8 transmission events are shown here. Green represents the rolling processing and pink represents traditional processing. The green and pink block represents the frame when the rolling reconstruction and traditional methods are equivalent.}
        \label{fig:rolling_processing}
\end{figure}

To build a super-resolution image, tracking is performed with the assumption of approximately linear displacement between frames. This assumption can only be satisfied if the frame rate used is sufficiently high, vessels are sufficiently straight and flow is sufficiently slow. These conditions are particularly likely to not hold in pathological tissue where vessels can be more tortuous and flow can be more erratic. Traditionally, plane-wave coherent compounding and \gls{opwc} is performed in a block-wise manner, i.e. for each block, a set of N-steered transmission-receptions events are performed, and the resulting data is combined into a single output frame. This has the disadvantage of reducing the frame rate by N compared to the pulse repetition frequency (where N is the number of row or column transmissions). This may not be significant for most applications, but 3-D \gls{rca} \gls{ulm} requires a higher acquisition rate. This work uses a technique for improving the effective frame rate of \gls{rca} processing, using a rolling reconstruction approach. A similar technique has been used with synthetic aperture beamforming~\cite{nikolov_recursive_1999} and this has been applied to the \gls{rca}~\cite{jensen_anatomic_2022}. A sliding compounding approach has also been developed and used for Doppler imaging~\cite{kang_high_2018}. Here, rather than simply compounding, the reconstruction combines \gls{rc-fmas} and \gls{asap}. In this case, intermediate output frames are reconstructed by combining events from neighbouring blocks. To remove potential spatial biases, the row-column transmissions are interleaved in a way that a row transmission is always followed by a column transmission. Each frame is reconstructed with 2N transmission-reception events, with the events picked increasing by two each time (one updated row and one updated column), see Figure~\ref{fig:rolling_processing}. This double update is necessary for both \gls{opwc} and \gls{rc-fmas} otherwise microbubbles would appear to travel along a 'stair-like' path. This increases the effective frame rate by a factor of N, not only making tracking easier but also allowing for the implementation of \gls{asap} without any loss in temporal resolution. Note that the acquisition frame rate is not in fact increased; rather, the reduction in the effective frame rate that is common to block-wise processing is eliminated. Smearing artefacts caused by fast-moving targets will still be present, as this is caused by targets moving during the block duration. In super-resolution, these artefacts are not a primary concern because the localisation should still place the microbubble close to a location it passed through as the smearing artefact will be aligned along the direction of travel of the microbubble. If the vessels are highly tortuous then the centroid of \gls{psf} might be outside the vessel. However, if in-painting is used in this same highly tortuous case, then the created path will be outside the vessel too. It should be noted that the increased computational cost of this method is not particularly high as the most computationally expensive element of the reconstruction is the producing the pre-combined \gls{das} volumes. The rolling processing can then use these same volumes to produce various inter-frame volumes, without the need to recompute the \gls{das} volumes. In this implementation which was not optimised, the \gls{rc-fmas} method took on average $\sim$3.2 s per frame and the \gls{rc-fmas} rolling method took $\sim$3.8 s for 11 frames (number of additional frames). The processing and data transfer times to and from the GPU are included here, which is typically an expensive step. It should be noted that the amount of data that needs to be stored does increase which can be problematic during longer acquisitions. In any final clinical implementation of this procedure, these volumes would not necessarily need to be stored, however.

\subsection{Acoustic Sub Aperture Processing}
\label{sec:asap}
\begin{figure} [htb]
    \centering
        \includesvg[inkscapelatex=false,width=1\columnwidth]{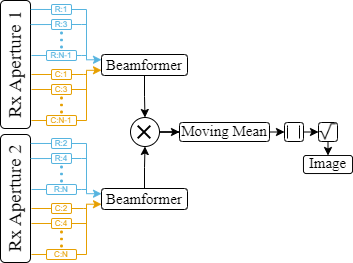}
        \caption{Diagram of the \gls{asap} image reconstruction procedure. The received data is split into two sub-apertures during post-processing. In this experiment, aperture one was every even row element (R) and column element (C), and aperture two consisted of the odd elements. All processing was then performed with two apertures separately, either using~\gls{opwc} or~\gls{rc-fmas} beamformers. The resulting two volumes were then combined via zero-lag cross-correlation and a moving average was taken to improve~\gls{snr}.}
        \label{fig:asap_pipeline}
\end{figure}

\gls{asap} is a reconstruction technique, developed by Stanziola et al.~\cite{stanziola_asap_2018} to improve power Doppler ultrasound imaging in the context of high frame rate, contrast-enhanced ultrasound. In this work,~\gls{asap} was implemented by splitting the received data into two separate apertures: the odd and evenly indexed channels. These apertures were then reconstructed independently with either~\gls{opwc} or~\gls{rc-fmas}. The zero-lag cross-correlation was subsequently taken between reconstructed volumes. For~\gls{asap} to have a benefit, multiple frames typically need to be averaged. This, however, reduces temporal resolution. Here, \gls{asap} was applied along with rolling compounding; if the length of the averaging window is chosen to match the number of compounding angles, the~\gls{snr} can be improved whilst not losing additional temporal resolution. Figure~\ref{fig:asap_pipeline} outlines the~\gls{asap} reconstruction pipeline. The square root of the resulting volumes was taken to remove dynamic range stretching. Two sub-apertures (odd and even elements) were used. Using interleaved sub-apertures increases the effective pitch of the transducer. Consequently, it is beneficial to start with a low pitch, which is why this technique is effective when using the \gls{rca}, particularly at low frequencies. These sub-apertures were chosen as the effective pitch of the transducer post~\gls{asap} would still be under twice the wavelength. As the field of view was only as wide as the probe, the grating lobes remained outside it even when \gls{asap} was used. As multiple frames need to be averaged to achieve any benefit from \gls{asap}, it could only be implemented in this case by integrating it with the rolling reconstruction, otherwise, major frame rate reduction would be experienced.

\begin{table}[ht]
\resizebox{\textwidth}{!}{%
\begin{tabular}{lccc}
\hline
                            & Cross-Tube          & Rabbit Kidney & Human Thyroid \\ \hline
Angle Range                 &  - 6$\degree$ $\rightarrow$ 6$\degree$&  - 6$\degree$ $\rightarrow$ 6$\degree$ &  - 6$\degree$ $\rightarrow$ 6$\degree$     \\           
Number of Angles            &  11 + 11            &  11 + 11      &  11 + 11      \\
Number of Cycles            &   1                 &  1            &  1     \\
Frequency                   &   3 MHz             &  3 MHz        &  3 MHz     \\
Frames per Batch            &   500               & 1350         &  2000     \\     
Frames per Batch (rolling)            &   5500               & 14850         &  22000     \\                         

Number of Batches               &   11                & 13            &  1      \\
SVD cutt-off                & 2\%                 & 7.5\%         &  15\%  \\
Noise Threshold (OPWC)      & 0.05                & 0.14          &  - \\ 
Noise Threshold (RC-FMAS)   & 0.03                & 0.07          & 0.08 \\
Minimum Extent              & 500 μm              & 400 μm        &  300 μm              \\
Minimum Solidity            & 0.4                 & 0.4           & 0.3               \\
Maximum Eccentricity        & 0.7                 & 0.7           & 0.8               \\
Auto-thresholding Step Size & 2.5$\times10^{-3}$  &  2.5$\times10^{-3}$ &  5$\times10^{-5}$ \\
Minimum Volume              & 0.5 μL     & 0.3 μL    & 0.1 μL                \\
Maximum Velocity            & 0.22 m/s   &   0.45 m/s        &  0.58 m/s       \\ \hline
\end{tabular}
}
\caption{Imaging and Processing Parameters.}
\label{tab:chapter:proccessing_paramters}
\end{table}
\subsection{Motion Correction}
\label{chap3:sec:moco}
Motion estimation was performed using non-clutter-suppressed reconstructed volumes beamformed prior to rolling reconstruction. As the dominant signal in this case came from the tissue, the motion of the microbubbles and blood would be minimal and thus ideally not have a large impact on the motion estimation. The reference frame was the most average frame, defined as the frame with the minimum difference to the average of all frames. All other frames were then corrected to this frame. Motion estimation was performed using the Matlab (The MathWorks, Inc., Natick, MA, USA) function \textit{imregdemons} using a GPU. The motion field was then used to correct the motion in the non-rolling reconstructed volumes directly. In the rolling reconstruction case, the motion field was estimated using the non-rolling reconstructed volumes and then cubic interpolation was used to generate inter-frame motion. This interpolated field was then applied to the rolling reconstructed volume. This was done as the motion estimation step is the slowest in the entire pipeline and thus performing it N more times would be computationally very expensive, both in terms of processing time and data storage as the motion field takes up a large amount of storage (three times more than beamformed volumes). This assumption of linear interframe motion will hold as long as the tissue motion is smooth and non-erratic which it was in both \text{in vivo} cases. For the \textit{imregdemons}  
three multi-resolution image pyramid levels were used and the iteration smoothing was set to one. As it is assumed that the \gls{rc-fmas} reconstruction will provide higher quality B-Mode images these were used to correct all data sets. This was to prevent the need for separate motion estimations.

\subsection{Ultrasound Localisation Microscopy }
\label{chap:3:sec:loc&track}
Prior to localisation, individual microbubbles need to be isolated. This was done using an auto-thresholding technique. Microbubbles were first isolated using an initial threshold, which was estimated by applying the Kittler-Illingworth method~\cite{kittler_minimum_1986} to the maximum intensity projection, in the time dimension, of the entire 4-D acquisition volume. This allowed for the automation of the initial thresholding step. Each continuous region above the threshold value was then analysed. If the region only contained one local maximum then it was deemed to be an isolated microbubble. If the region has multiple regional maxima, then the eccentricity and solidity of the region were evaluated, using the Matlab function `\textit{regionprops3}'.  Isolated microbubbles were assumed to be roughly spherical and continuous. If the region had high eccentricity or low solidity, it was deemed to not be a single microbubble and the threshold was raised to try and isolate the microbubble. Table~\ref{tab:chapter:proccessing_paramters} shows all localisation parameters used, unless otherwise specified all parameters were determined via manual visual assessment of a few frames. This technique was repeated until each region was deemed to contain only one isolated microbubble. If the volume and extent of the microbubble were too small, then the region was deemed to be noise and removed. Noise regions that have passed through initial thresholding will typically have a much smaller spatial extent than microbubbles. Weighted centroiding was used to localise the microbubble within the field of view~\cite{siepmann_imaging_2011}. 
After localisation, tracking was used to remove noise/artefact signals that may have passed through previous processing steps and to populate inter-frame elements of the vessels. Microbubbles were assumed to follow a linear path between frames. The validity of this assumption is dependent on the effective frame rate and thus implementing a rolling reconstruction technique will reduce inaccuracies caused by non-linear displacement.  For tracking, an in-house Kalman filtering algorithm was used~\cite{yan_super-resolution_2022, yan_transthoracic_2023}. In this technique, microbubble pairing was achieved by calculating the total minimum cost through graph-based assignment, where the cost is defined by the ratio of the microbubble intensity difference to the probability obtained from the Kalman motion model. The pairing was further refined by independently discarding microbubble pairs that had too great a deviation from the motion model, or too large an intensity change. The search window used for determining microbubble pairings was calculated using the maximum velocity relevant to that particular organ and model~\cite{chu_morphology_2011,turgut_maximum_2009}.
All tracking parameters used can be found in Table~\ref{tab:chapter:proccessing_paramters}, unless otherwise specified all parameters were chosen via visual inspection of results from processing a few frames. 

\subsection{Microbubbles}
The contrast agents used for the cross tube and rabbit experiments were in-house prepared perfluorobutane microbubbles. They were prepared by dissolving 1,2-Dipalmitoyl-sn-glycero-3-phosphatidylcholine, 1,2-dipalmitoyl-sn-glycero-3-phosphatidyleth-anolamine-polyethylene glycol 2000 and 1,2-dipalmitoyl-3-trimethylammonium propane in water with a molar ratio of 65:5:30 and a total lipid concentration of 0.75, 1.5 and 3 mg/mL respectively. This resulted in a solution with 15\% propylene, 5\% glycerol and 80\% saline.  Perfluorobutane was then injected into a 2 ml container containing 1.5 ml of the prepared solution just before use. The microbubbles were activated by agitation in a shaker for 60 seconds. Such microbubbles have previously been counted and found to have a concentration of $\mathrm{5\times10^{9} \: microbubbles/mL}$ ~\cite{leow_flow_2015}. For the human experiments, SonoVue\textsuperscript{\textregistered} (Bracco, Milan, Italy) was used as the ultrasound contrast agent. SonoVue was not available for the cross-tube and rabbit experiments as it needed to be supplied to a clinician. 
\subsection{{\it{In Vitro:}} Cross Tube}

\begin{figure} [htb]
    \centering
        \includegraphics[width=1\linewidth]{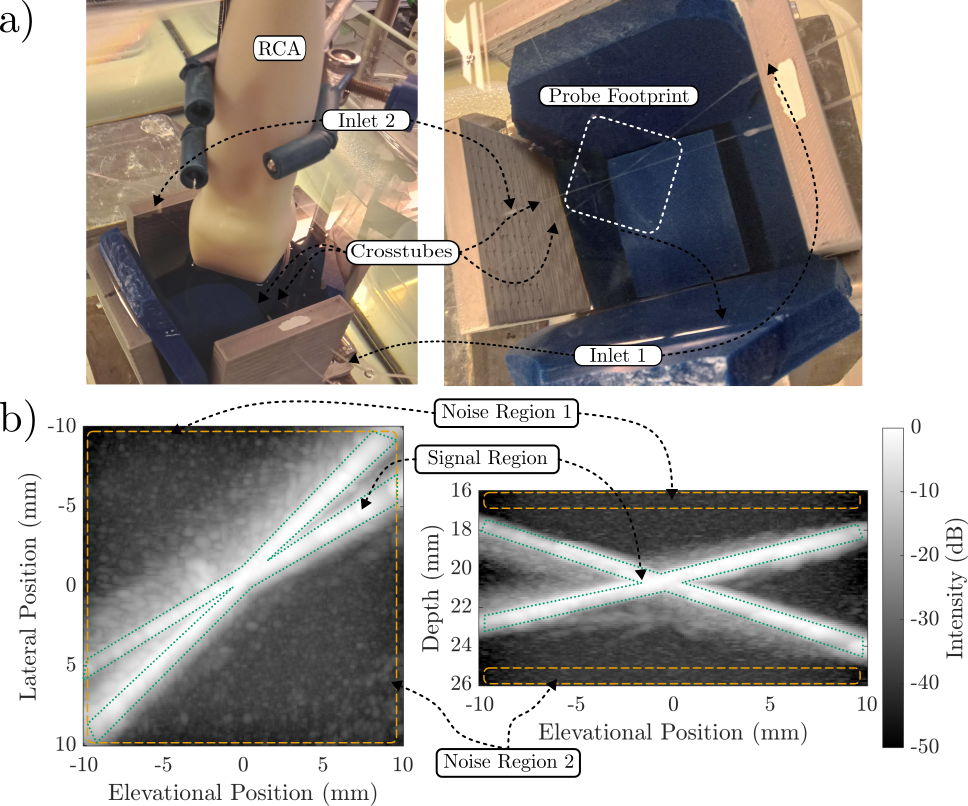}
        \caption{a) Shows the cross-tube experimental setup. b) Shows Lateral/Elevational \gls{mip} and Axial/Elevational \gls{mip} using \gls{rc-fmas} with rolling \gls{asap} image reconstruction. Intensity is based on the temporal average over all frames. For \gls{snr} calculations, the green dotted region corresponds to the signal, while the two orange dashed regions correspond to noise regions that will be averaged. These noise regions were chosen so that they do not contain artefacts caused by the various image reconstruction techniques.}
        \label{fig:cross_tube_setup}
        
\end{figure}

Two Hemophan cellulose tubes (Membrana, 3M, Wuppertal, Germany) with 200 µm ± 15 µm outer diameter and wall thickness of 8 ± 1 µm were placed crossing over each other in the field of view of the \gls{rca}. To minimise any potential directional bias, the tubes were aligned diagonally such that neither aligned directly with either the rows or the columns. The cross tubes could not physically touch so centres were separated by $\sim$233 ± 20 µm; thus the edges of the tubes were separated by less than the half-wavelength distance of 257 µm.  The distance was estimated by drawing two straight lines through both tubes in 3-D space using the temporal maximum projection image and calculating the distance between the two lines. For contrast agents, the homemade microbubbles were diluted to a concentration of approximately $\mathrm{2\times10^{4} \: μB/mL}$. The cellulose tubes were connected via silicon tubing to a syringe pump (Harvard Apparatus, Holliston, MA, USA) which provided an average flow of velocity 0.11 m/s inside the cross tubes. Figure~\ref{fig:cross_tube_setup} shows the cross-tube setup. 11 row and 11 column transmissions were used. The angle range chosen was from -6$\degree$ to +6$\degree$. These values were chosen as they were found to give the best image quality whilst maintaining a high frame rate. They were chosen based on the optimisation performed in prior work~\cite{hansenshearer_rc_fmas}. As the cross-tube was static, a low \gls{svd} cut-off threshold (2\% of batch frames) was sufficient to remove clutter. Rolling \gls{opwc} and \gls{rc-fmas} combined with \gls{asap} were used to reconstruct the volumes of the image. Auto-thresholding and Kalman filtering were then used for localisation and tracking.  15 acquisitions, each with 500 frames, were combined to form the final image, resulting in a total active acquisition time of 17 s. Time was needed between each acquisition to save the data.

\subsection{{\it{In Vivo:}} Rabbit Kidney}
A specific-pathogen-free New Zealand White rabbit (male, HS-DIF strain, age 13 weeks, weight 2.4 kg, Envigo, UK) was sedated with acepromazine (0.5 mg/kg, i.m.) and anaesthetised with medetomidine (Domitor, 0.25 mL/kg, i.m.) and ketamine (Narketan, 0.15 mL/kg, i.m.). The experiment was authorised by the Animal Welfare and Ethical Review Body of Imperial College London and conformed with the Animals (Scientific Procedures) Act 1986. After being anaesthetised the fur overlying the kidney was shaved and imaged with the \gls{rca}, non-invasively. Homemade microbubbles were introduced into the circulation via bolus injection into the marginal ear vein followed by a saline flush. Two injections were performed one $\sim$20 before the first acquisition and one $\sim$20 before the 8th acquisition. The rabbit was kept static and the probe was held using a clamp once the correct position had been found. 11 row and 11 column transmissions were used which resulted in a frame rate of 441 fps. The angle range chosen was from -6$\degree$ to +6$\degree$. 13 acquisitions each with 1350 frames were combined to produce the final image. The total imaging time was 39.8 s. Motion correction was applied across all batches and \gls{svd} was calculated separately on each batch and angle with a \gls{svd} cut-off of 7.5\% of batch frames. 0.3 ml of microbubbles were injected in total, in three injections, throughout the acquisitions.  The maximum imaging depth was 3 cm.

\subsection{{\it{In Vivo:}} Human Thyroid}
Images were acquired of the thyroid of a male volunteer from an arrhythmia outpatient clinic group participating in a prospective clinical cohort study investigating the additive diagnostic benefit of high frame rate contrast-enhanced ultrasound. The study was reviewed and approved by the London-Bromley Research Ethics Committee and the Health Research Authority (IRAS Project ID 144257, REC reference 14/LO/0360) and is ongoing. The patient was informed and gave written consent to participate in the trial. The patient was given 2 ml of Sonovue via bolus IV injection, followed by a saline flush. 20~s after injection, data were acquired for 12~s using at a frame rate of 441 fps. The duration was limited by the amount of data that could be acquired by the Verasonics system in one acquisition before it ran out of channel RAM. An \gls{svd} cut-off of 15\% of all frames was used. Multiple acquisitions were not possible as there would be too much motion between acquisitions. The maximum imaging depth was 3.5 cm. 11 row and 11 column transmissions were used. The angle range chosen was from -6$\degree$ to +6$\degree$. 

\subsection{Image Quality Assessment}
Removing non-main lobe signals before performing localisation is commonly used in super-resolution imaging. The improvement offered by \gls{rc-fmas} with regards to `secondary' lobe reduction has already been investigated~\cite{hansenshearer_rc_fmas}. Here, the noise reduction of the various techniques was analysed in two ways. First, the traditional \gls{snr} was calculated:

\begin{equation}
SNR =  \left [ \frac{μ_{signal}}{μ_{noise}}  \right ]_{dB},
\end{equation}
where $μ_{signal}$ is the mean power in the signal region defined as the area within cross-tubes denoted by green dotted lines in Figure~\ref{fig:cross_tube_setup} and $μ_{noise}$ is the mean power in a noise region above and below the cross tube shown by the yellow dashed box in Figure~\ref{fig:cross_tube_setup}. The \gls{snr} was calculated for each frame and then averaged. When investigating image quality for super-resolution, the variation in the noise is sometimes more important than the mean. If the noise is flat (low variance) then it is easier to remove than if there are large noise peaks which could be misinterpreted as a true microbubble signal. Typically, prior to microbubble isolation, noise is thresholded out. The method for determining this threshold can be challenging but where possible it is preferable to pick a threshold at least 3σ above the noise mean. For this purpose, an alternative definition of the \gls{snr} was used: 

\begin{equation}
SNR_{3σ} =  \left [ \frac{μ_{signal}}{μ_{noise}+3σ_{noise}}  \right ]_{dB},
\end{equation}
where $σ_{noise}$ is the standard deviation of the noise region. 

To evaluate the resolution the \gls{fsc} was used. It is the 3-D implementation of the Fourier Ring Correlation~\cite{nieuwenhuizen_measuring_2013,banterle_fourier_2013,hingot_measuring_2021}. A 5σ threshold was used to estimate the resolution achieved, which identifies structures systematically above the noise correlation~\cite{van_heel_fourier_2005}. It is not expected that the \gls{rc-fmas} method will improve the resolution significantly; rather, it will reduce the number of false localisations. {\it{In vivo}} there is no ground truth available, and there is no explicit ground truth in the {\it{in vitro}} case. In the latter case, we can estimate the number of `false' localisations by defining them as any localisation outside the cross-tube regions. `False' localisations are expressed as a percentage of total number of localisations. The tube region was set by setting a centre line through both cross tubes. Any microbubble within 143 µm (radius of tube plus manufacturing error, localisation error and centre line uncertainty) of these lines was determined to be within the tube region, see the green dotted region in Figure~\ref{fig:cross_tube_setup}.

\section{Results}
\subsection{{\it{In Vitro:}} Cross Tube}
\begin{figure} [htb]
    \centering
        \includegraphics[width=0.8\linewidth]{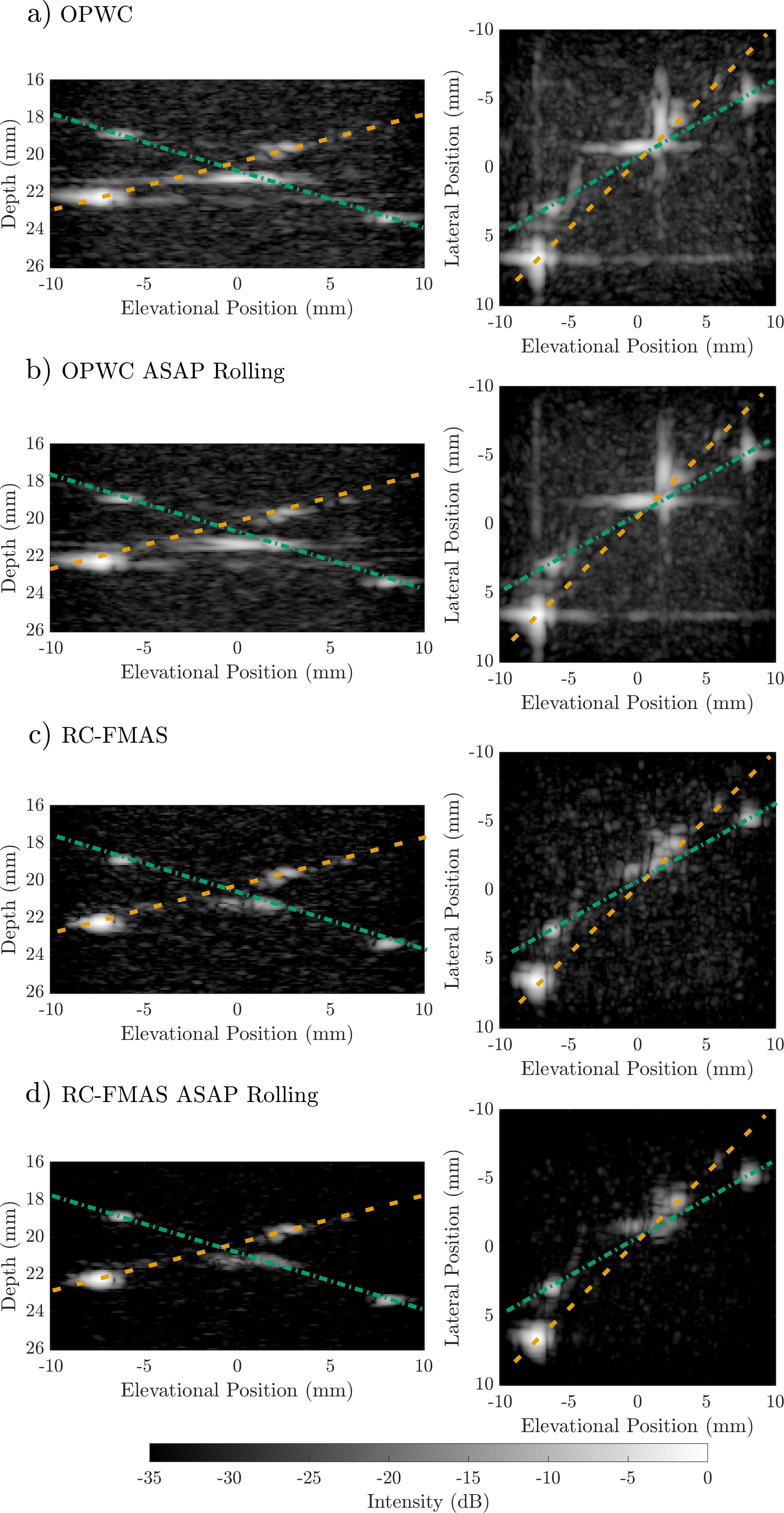}
        \caption{Shows a \gls{mip} of a single frame used of the cross tube experiment using a) \gls{opwc}, b) \gls{opwc} with \gls{asap} rolling reconstruction, c) \gls{rc-fmas} and d) \gls{rc-fmas} with  \gls{asap} rolling reconstruction. The left column is the \gls{mip} along the lateral direction and the right column is the \gls{mip} along the axial direction. The green (dashed and dotted) and orange (dashed) lines represent the approximate locations of the two microtubes.}
        \label{fig:rcfmas_cross_tube_frame}
        
\end{figure}

Figure~\ref{fig:rcfmas_cross_tube_frame} shows a single frame reconstructed with~\gls{opwc} with and without~\gls{asap}, and~\gls{rc-fmas} with and without~\gls{asap}. By averaging all the frames \gls{opwc} was found to have a~\gls{snr} of 18.4 ± 0.7 dB, without ~\gls{asap} and 18.0 ± 0.7 dB with ~\gls{asap}. ~\gls{rc-fmas} was found to have a~\gls{snr} of 24.8 ± 0.9 dB, without ~\gls{asap} and 25.7 ± 0.9 dB with ~\gls{asap}. When looking at \gls{snr}\textsubscript{3σ} very little improvement was observed when performing \gls{asap}, with \gls{opwc} the \gls{snr}\textsubscript{3σ} was 12.7 ± 0.7 dB with \gls{asap} and 12.7 ± 0.7 dB without it. However, in \gls{rc-fmas} case a more significant improvement is seen with the \gls{snr}\textsubscript{3σ} going from 16.9 ± 0.9 dB without \gls{asap} to 19.3 ± 0.9 dB with it. It is believed that this performance is due to the noise characteristic changing when performing \gls{rc-fmas} compared to simply compounding volumes as with \gls{opwc}. Each \gls{snr} and \gls{snr}\textsubscript{3σ} value is calculated independently for each data-set used, the final number quoted is the mean result and the error is the standard error of the mean.

\begin{figure*} [ht]
    \centering
    \includegraphics[width=0.8\textwidth]{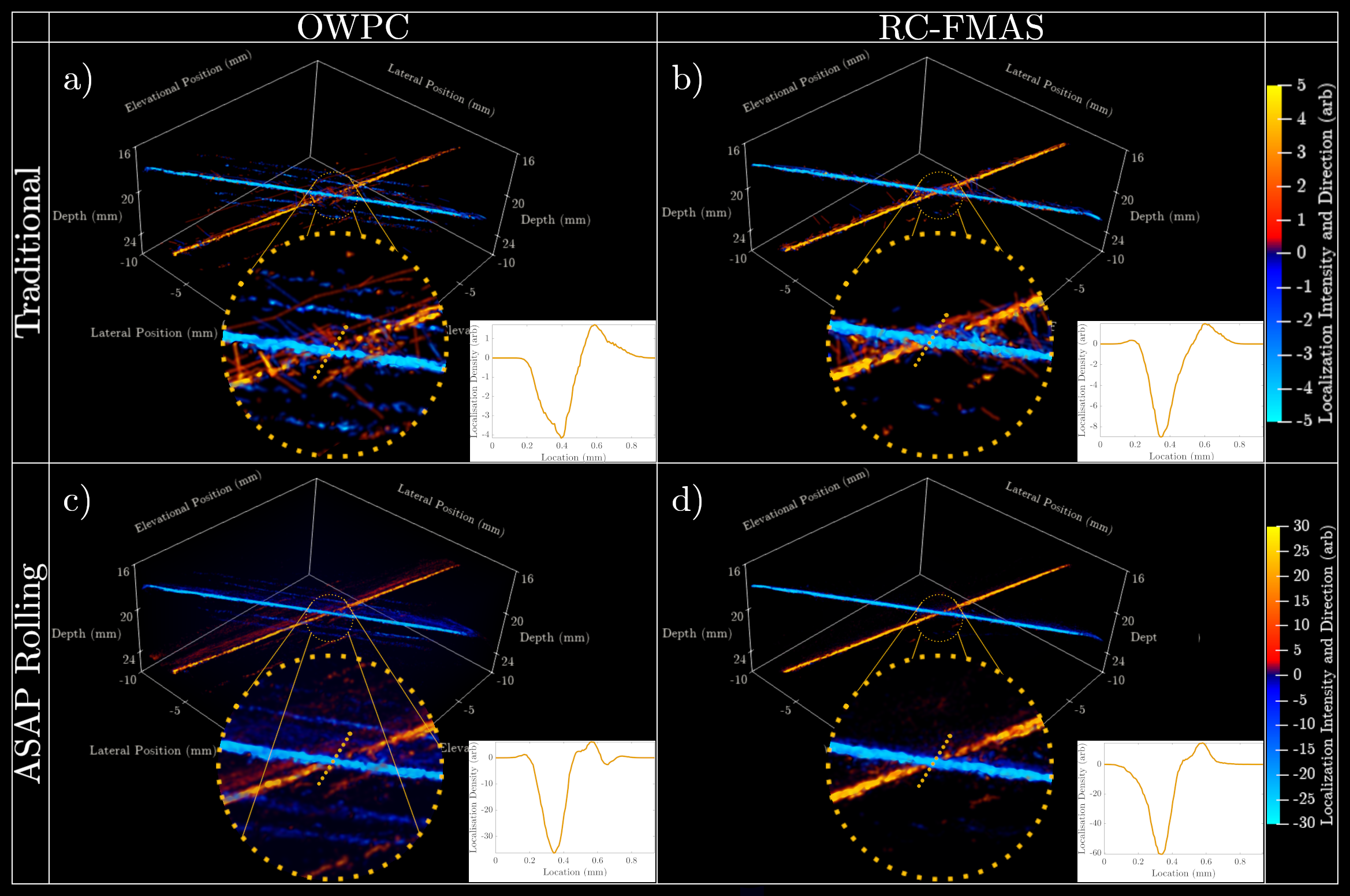}
    \caption{Shows super-resolution images cross tube using all 4 techniques. Red represents flow upward, the blue represents flow downward. This is done to distinguish the tracks from one another. A zoom-in of the crossing point is shown in the yellow dashed circles. A line profile (dotted yellow line) through the two tubes at their closest point is shown for each technique. }   
    \label{fig:sr_cross_tube}
\end{figure*}

Super-resolution images of two cross tubes were generated using both \gls{opwc} and \gls{rc-fmas}, again with and without~\gls{asap}, see Figure~\ref{fig:sr_cross_tube}. When \gls{asap} rolling reconstruction was used, both \gls{opwc} and \gls{rc-fmas} managed to separate the two vessels, as shown by the line profiles in Figure~\ref{fig:sr_cross_tube}.
More important for the context of this study is the number of false localisations. From Figure~\ref{fig:sr_cross_tube} it can be seen that the number of `false' localisations outside of the cross tube region is minimised when using \gls{rc-fmas} and when using \gls{asap} rolling reconstruction. The \gls{asap} rolling compounding method decreases the number of noise events and false tracks caused by insufficient frame rate. The \gls{rc-fmas} method reduces the number of false localisations caused by the `secondary' lobe artefact. The straight tubes appearing parallel to the actual cross tubes are caused by these `secondary' lobes being localised and tracked. As the quality of the final super-resolution image is dependent on the thresholds chosen, it is challenging to compare techniques quantitatively. Specifically, the threshold chosen can greatly change the quality of the final image. When using \gls{asap} rolling reconstruction the number of localisations was 231,807 when using \gls{opwc} and 238,984 when using \gls{rc-fmas}, compared to 24,786 and 27,445 when using just \gls{opwc}  and \gls{rc-fmas} respectively. As the number of localisations was higher for \gls{rc-fmas} methods and yet fewer `false' localisations (localisations outside tube region) were detected, the thresholds chosen were appropriate for comparison. The \gls{fsc} 5σ thresholds and `false' localisation percentages, along with the two \gls{snr} estimations can be found in Table~\ref{tab:cross_tube_analysis}. 
\begin{table}[ht]
\caption{Cross Tube Analysis}
\label{tab:cross_tube_analysis}
\resizebox{\textwidth}{!}{%
\begin{tabular}{lllll}

\hline
                  & SNR  [dB]     & SNR\textsubscript{3σ} [dB] & FSC\textsubscript{5σ} & \makecell{False \\ Localisations} \\ \hline
OPWC                 & \multicolumn{1}{c}{18.4±0.7} & \multicolumn{1}{c}{12.3±0.7} & \multicolumn{1}{c}{126 μm} &  \multicolumn{1}{c}{6444 (26\%)}  \\
OPWC ASAP Rolling    & \multicolumn{1}{c}{18.0±0.7} & \multicolumn{1}{c}{12.7±0.7} & \multicolumn{1}{c}{114 μm} &  \multicolumn{1}{c}{60270 (26\%)} \\
RC-FMAS              & \multicolumn{1}{c}{24.8±0.9} & \multicolumn{1}{c}{16.9±0.9} & \multicolumn{1}{c}{118 μm} &  \multicolumn{1}{c}{4666 (17\%)}  \\
RC-FMAS ASAP Rolling & \multicolumn{1}{c}{25.7±0.9} & \multicolumn{1}{c}{19.3±0.9} & \multicolumn{1}{c}{107 μm} &  \multicolumn{1}{c}{35848 (15\%)} \\ \hline

\end{tabular}
}
\end{table}

\subsection{{\it{In Vivo:}} Rabbit Kidney}
\begin{figure*} [t]
    \centering
    \includegraphics[width=1\textwidth]{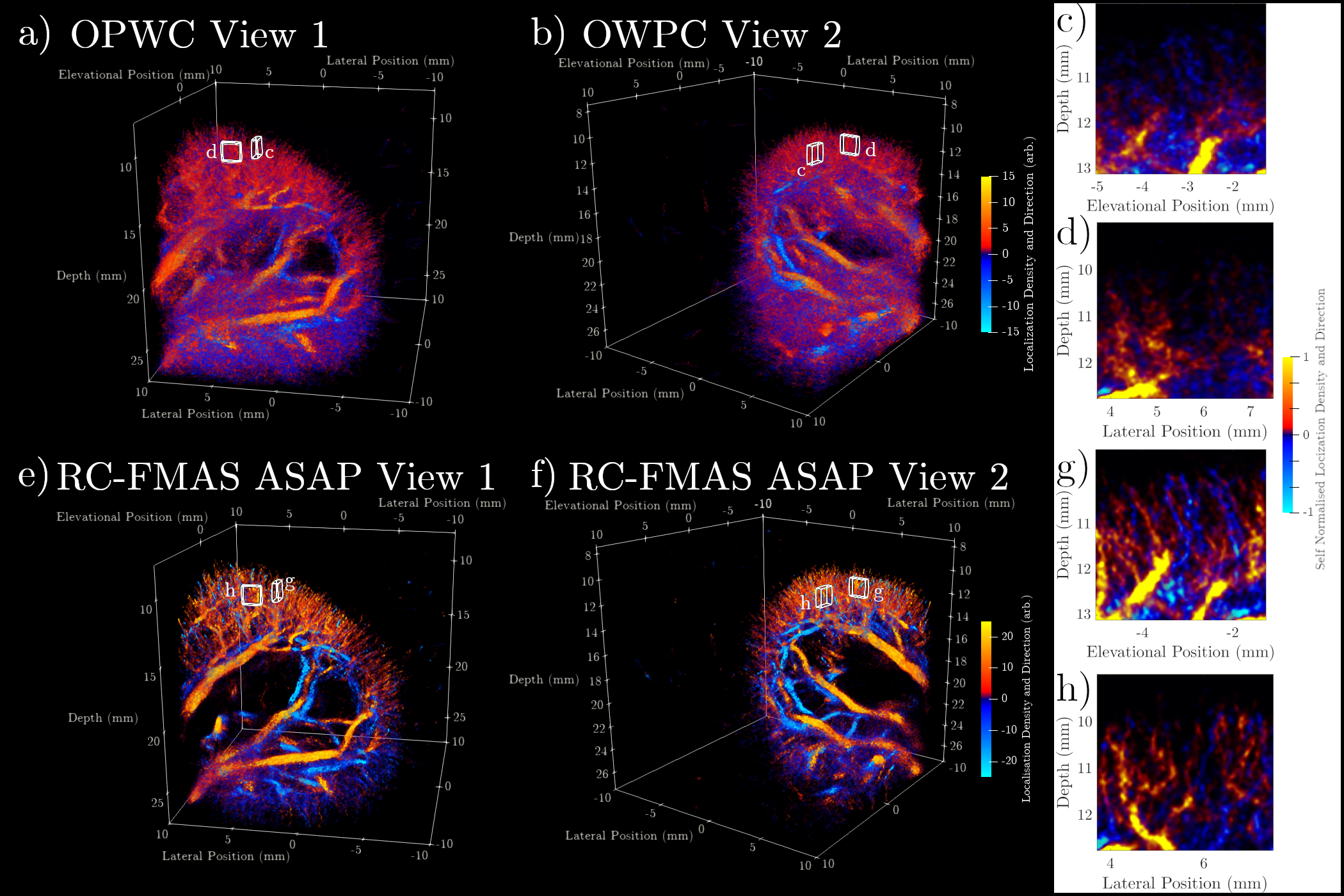}
    \caption{a) and b) show projection images of the non-invasively imaged rabbit kidney using \gls{opwc} from 2 different views. The intensity shows the number of localisations and the colour represents flow direction: red is up, blue is down. c) and d) are projections taken from the 3-D volume shown in a) and b). e) and f) show the same views as a) and b) but reconstructed with the \gls{rc-fmas} technique along with \gls{asap} rolling compounding. g) and h) are projections taken from the 3D volume shown in e) and f). The locations of the projections are denoted by the white boxes displayed in the other slices. c) and g) show projections along the lateral direction and d) and h) show projections along the elevational direction. For 3-D visualisation, an intensity opacity map was used. The map for \gls{rc-fmas} was linear. To visualise the \gls{opwc} case, non-linear opacity and colour map were needed. These were needed otherwise no structures could be visualised. The maps were generated using manual inspection. Figures c), d), g) and h) are all self-normalised, so colours cannot be compared to the 3-D visualisation. } 
    \label{fig:sr_rabbit}
\end{figure*}

As there is no prior knowledge available when imaging {\it{in vivo}}, the techniques cannot be compared directly. The {\it{in vivo}} processing was conducted using \gls{rc-fmas} with \gls{asap} rolling reconstruction and then these results were compared with those obtained using the baseline \gls{opwc} algorithm. Figure~\ref{fig:sr_rabbit} shows the resulting super-resolution image obtained with both techniques. The \gls{fsc} 5σ threshold was 313 μm for the \gls{opwc} and 232 μm for \gls{rc-fmas} with \gls{asap} rolling reconstruction. 

\subsection{{\it{In Vivo:}} Human Thyroid}
\begin{figure*} [t]
    \centering
    \includegraphics[width=1\textwidth]{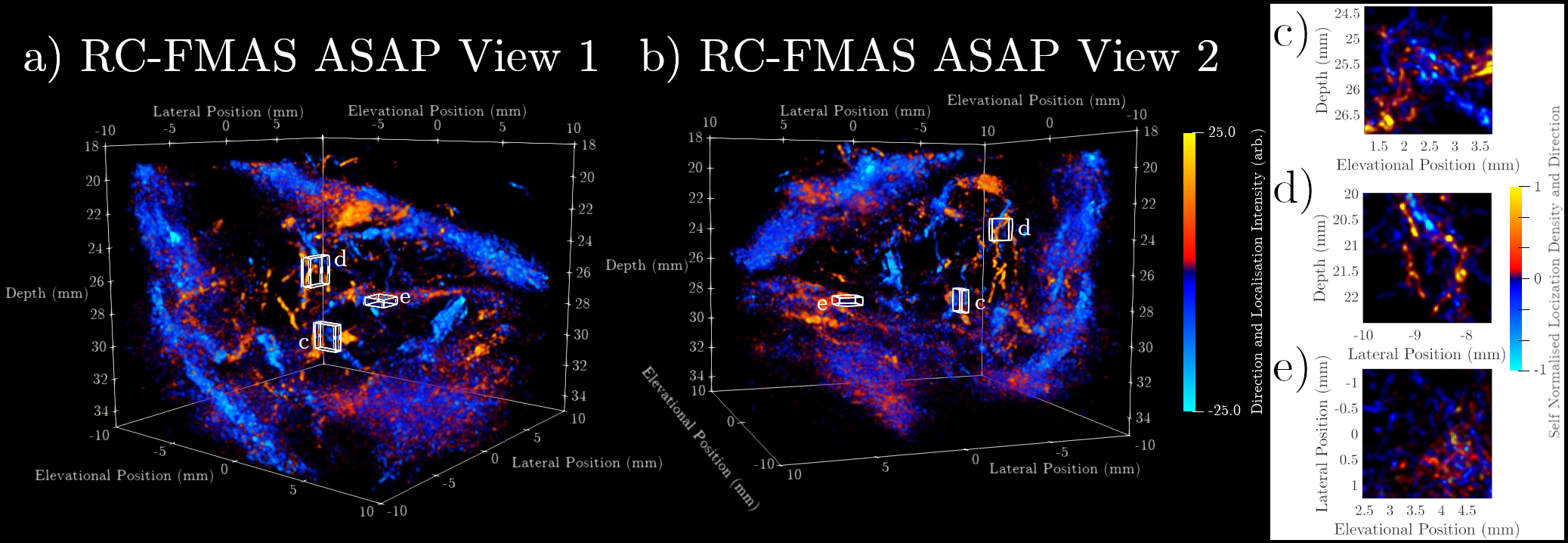}
    \caption{a) and b) show projection images of the human thyroid from 2 different views using the \gls{rc-fmas} method. Intensity shows the number of localisations and colour represents flow direction: red is up, blue is down. c), d) and e) are slices taken from the 3D volume. The location of the slices is denoted by the 3 white boxes displayed in a) and b). the slices are orthogonal to one other.  Figures c), d) and e) are all self-normalised, so colours cannot be compared to the 3-D visualisation. } 
    \label{fig:sr_thyroid}
\end{figure*}

Figure~\ref{fig:sr_thyroid} shows a human thyroid imaged with the \gls{rca} and processed according to the full pipeline outlined in Figure~\ref{fig:processing_pipeline}, using \gls{rc-fmas} with \gls{asap} rolling reconstruction.

 As imaging time was limited and multiple acquisitions were not possible, the final super-resolution image does not have a fully populated vascular network. This is not a limitation of the technique but of the clinical setting. By holding the probe in the same location and optimising the infusion of Sonovue, it should be possible to image the patient for a few minutes without disruptions to the clinical workflow. Longer acquisitions would require a more involved solution and might need mechanical fixing of the probe. Note that the vascular network was populated more thoroughly in the rabbit kidney (Figure~~\ref{fig:sr_rabbit}) with only 38.9~s of acquisition. This length of acquisition would be achievable in a clinical setting. As this is a 3D imaging technique, motion correction is easier, as out-of-plane (volume) motion has less impact. Thus even if the probe drifts due to patient or clinician movement, it can be compensated more easily than in the 2-D case. The \gls{fsc} 5σ threshold for the thyroid image was 409 μm. The wavelength in this experiment was again 513 μm.

\section{Discussion}
In this work, 3-D super-resolution imaging with a large field of view was achieved with an \gls{rca}, \textit{in vitro} and \textit{in vivo} using the recent beamforming technique: \gls{rc-fmas}. \textit{In vitro}, a cross tube was used to evaluate the resolution achievable with the technique and to demonstrate the increase in precision possible by the suppression of `secondary' lobe artefacts present when operating a \gls{rca} in the ultrafast regime. A rabbit kidney was then imaged non-invasively to produce a super-resolution image using the \gls{opwc} and our proposed algorithms. Finally, 3-D \gls{ulm} was demonstrated in a human using an \gls{rca} for the first time by imaging the thyroid. 

The \gls{rca} has a large aperture making it highly suitable for super-resolution imaging as the field of view can be larger than a matrix array probe with comparable channel count using plane waves compounding. If a lens were to be integrated to allow off-axis diverging waves then the field of view of \gls{rca} could be comparable to that of a linear probe~\cite{bouzari_curvilinear_2017}. The major advantage of \gls{rc-fmas} is the ability to achieve ultrafast ultrasound imaging whilst keeping artefact level low. This applies particularly to specific artefacts such as the `secondary' lobes, which can be easily mistaken for microbubbles. By implementing \gls{asap} with rolling compounding, the noise and noise variance in the image were reduced whilst the effective frame rate was increased. The reduction in noise variance is particularly useful in the context of super-resolution as low-variance noise is easier to be removed via thresholding than high-variance noise. This allowed easier tracking of microbubbles, aiding in the generation of the final super-resolution images. 

In the cross-tube experiment, a small number of the `false' localisations will be caused by true air bubbles in the water tank but the majority come from either `secondary' lobes being detected as microbubbles or from events being tracked incorrectly. Such false negative events will be present in all techniques, so the percentage of false localisations can be used to compare them. In Figure~\ref{fig:sr_cross_tube} it can be seen in the \gls{opwc} cases that a series of `false' tubes are located parallel to the main tubes. These occurred as the side lobes were consistently detected as primary microbubbles. They can be avoided by raising the initial threshold level, but that would reduce the recall of the system as weaker true microbubble signals would be removed. To achieve super-resolution, the vessels being imaged need to be highly saturated with microbubbles throughout the imaging time. In a clinical setting, it is typically not possible to image the patient for extended periods (more than a few minutes) so any reduction in the number of localisations per unit time can be highly detrimental. All the metrics indicated that the \gls{rc-fmas} with \gls{asap} rolling reconstruction produced the best quality cross-tube result, and that agrees with the qualitative assessment. It should also be noted that the majority of `false' localisations for the \gls{rc-fmas} cases are still close to the main tube. Reverberations in the water tank, slight tube motion and interference caused by incompletely isolated microbubbles are likely to be the main causes.  Increasing the distance from the tube centre-line over which localisations are considered true decreases the number of `false' localisations significantly in the \gls{rc-fmas} cases. In the \gls{opwc} cases, however, there are a large number of `false' localisation ($\sim$10 \% of all localisations) caused by the distant `secondary' lobes. From this, it can be inferred that the \gls{rc-fmas} will be able to generate more correct localisation in an \textit{in vivo} context, where choosing the localisation threshold can be more challenging, or alternatively more vessels can be populated within a shorter acquisition window. 

By implementing \gls{asap} rolling reconstruction, the effective frame rate can be increased to over 4000 Hz. This allows much easier tracking as the displacement of microbubbles is lower. It particularly aids the tracking of faster-flowing microbubbles. It also leads to a more accurate reconstruction of the vessels. To produce the final super-resolution image, tracked events are linked and an assumption about the inter-frame microbubble location is made. It is typically assumed that the displacement is linear, which might not always be the case. By decreasing the displacement between frames the errors introduced by this assumption can be reduced and a more accurate picture of the vessel can be built. In this work two volumes are updated in the rolling reconstruction technique, increasing the effective frame rate by a factor of N. These very high frame rates are not necessarily needed, for some applications. In these situations, a reduced frame rate could be used which would reduce the computation processing time and quantity of data that needs to be handled.

The \gls{asap} operation allows for an increased \gls{snr} when combined with the \gls{rc-fmas} algorithm. This is due to the \gls{rc-fmas} technique changing the noise characteristics: by only averaging frames used in rolling reconstruction, the noise and, more importantly for microbubble isolation, the noise variance is reduced. In future work, the averaging ensemble length should be investigated. By averaging over more frames, the noise could be reduced further, but at the cost of temporal resolution. For some applications reduced temporal resolution can be accepted. Using the 5σ threshold for the rabbit examples, sub-half-wavelength resolution was not achieved using just \gls{opwc} but by implementing \gls{rc-fmas} with rolling \gls{asap}, structures smaller than the half-wavelength were visualised above the noise background.

Figure~\ref{fig:sr_rabbit} shows improvement in the quality of super-resolution images when using the \gls{rc-fmas} method and \gls{asap} rolling reconstruction. Due to interference from various artefacts and the high noise levels present, tracking and localisation were not able to reconstruct fully the vascular tree. Figure~\ref{fig:sr_rabbit} c), d), g) and h) show projections through small subsections of the images, focusing on the microvasculature. In the g) and h), vascular tracks are visible that cannot be seen well in the \gls{opwc} case.

For the thyroid example, using the 5σ threshold for resolution estimation, sub-half-wavelength resolution cannot be claimed. This may be due to the low saturation of localisation events. When a dataset is undersampled or 
has low saturation the \gls{fsc} metric performs less well~\cite{hingot_measuring_2021}. In future work, longer acquisition durations will be used. This will be achieved by streamlining the data-saving process and by increasing the period over which microbubbles are present in the bloodstream. Due to specific clinical constraints, only one bolus injection was possible; this leads to a high peak concentration of microbubbles followed by a rapid decay, giving only a short time window having the optimum concentration ($\sim$ 30~s). The time when this optimum window occurs is patient-dependent so timing is imprecise. For future acquisitions, longer boluses or infusion injections will be used to maintain useful microbubble concentrations for longer.
In future work, improved saving techniques and microbubble injection procedures will be implemented to enable longer acquisition duration which will allow for more complete images of the vascular network. This work has managed to show that the \gls{rca} is able to produce super-resolution images non-invasively \textit{in vivo}, but that more work is needed to streamline the process in the clinic. Investigations into alternative localisation techniques and parameter selection would be useful. 
 
\section{Conclusion}
Here an \gls{rca} coherence-based beamforming technique was implemented to generate large field-of-view, super-resolution images with only 256 independent channels within clinically achievable time-frames. \gls{rc-fmas} was shown to decrease the number of false localisations by reducing the size of the `secondary' lobes, thereby reducing the number of false vessels being produced. The technique was combined with \gls{asap} rolling compounding to improve the \gls{snr} and effective temporal resolution. This work showed the feasibility of producing super-resolution images transcutaneously {\it{in vivo}} in rabbits. It is believed by the authors that it is the first demonstration of the feasibility of 3-D ultrasound localisation microscopy in a human patient using an \gls{rca} probe.
 
\printbibliography 

@article{denarie_coherent_2013,
	title = {Coherent Plane Wave Compounding for Very High Frame Rate Ultrasonography of Rapidly Moving Targets},
	volume = {32},
	issn = {1558-254X},
	doi = {10.1109/TMI.2013.2255310},
	pages = {1265--1276},
	number = {7},
	journaltitle = {{IEEE} Transactions on Medical Imaging},
	author = {Denarie, Bastien and Tangen, Thor Andreas and Ekroll, Ingvild Kinn and Rolim, Natale and Torp, Hans and Bjåstad, Tore and Lovstakken, Lasse},
	date = {2013-07},
}

@article{ceroici_fast_2019,
	title = {Fast Orthogonal Row-Column Electronic Scanning ({FORCES}) Experiments and Comparisons},
	issn = {1525-8955},
	doi = {10.1109/TUFFC.2019.2906599},
	journaltitle = {{IEEE} transactions on ultrasonics, ferroelectrics, and frequency control},
	shortjournal = {{IEEE} Trans Ultrason Ferroelectr Freq Control},
	author = {Ceroici, Chris and Lathamm, Katherine and Greenlay, Ben and Brown, Jeremy A. and Zemp, Roger},
	date = {2019-03-20},
	pmid = {30908213},
}

@article{sauvage_large_2018,
	title = {A large aperture row column addressed probe for in vivo 4D ultrafast doppler ultrasound imaging},
	volume = {63},
	issn = {0031-9155},
	url = {https://dx.doi.org/10.1088/1361-6560/aae427},
	doi = {10.1088/1361-6560/aae427},
	pages = {215012},
	number = {21},
	journaltitle = {Physics in Medicine \& Biology},
	shortjournal = {Phys. Med. Biol.},
	author = {Sauvage, J. and Flesch, M. and Férin, G. and Nguyen-Dinh, A. and Porée, J. and Tanter, M. and Pernot, M. and Deffieux, T.},
	urldate = {2023-08-01},
	date = {2018-10},
	langid = {english},
}

@article{van_heel_fourier_2005,
	title = {Fourier shell correlation threshold criteria},
	volume = {151},
	issn = {1047-8477},
	url = {https://www.sciencedirect.com/science/article/pii/S1047847705001292},
	doi = {10.1016/j.jsb.2005.05.009},
	pages = {250--262},
	number = {3},
	journaltitle = {Journal of Structural Biology},
	shortjournal = {Journal of Structural Biology},
	author = {van Heel, Marin and Schatz, Michael},
	urldate = {2023-08-01},
	date = {2005-09-01},
	langid = {english},
}

@article{banterle_fourier_2013,
	title = {Fourier ring correlation as a resolution criterion for super-resolution microscopy},
	volume = {183},
	issn = {1047-8477},
	url = {https://www.sciencedirect.com/science/article/pii/S1047847713001184},
	doi = {10.1016/j.jsb.2013.05.004},
	pages = {363--367},
	number = {3},
	journaltitle = {Journal of Structural Biology},
	shortjournal = {Journal of Structural Biology},
	author = {Banterle, Niccolò and Bui, Khanh Huy and Lemke, Edward A. and Beck, Martin},
	urldate = {2023-08-01},
	date = {2013-09-01},
	langid = {english},
}

@article{nieuwenhuizen_measuring_2013,
	title = {Measuring image resolution in optical nanoscopy},
	volume = {10},
	issn = {1548-7105},
	doi = {10.1038/nmeth.2448},
	pages = {557--562},
	number = {6},
	journaltitle = {Nature Methods},
	shortjournal = {Nat Methods},
	author = {Nieuwenhuizen, Robert P. J. and Lidke, Keith A. and Bates, Mark and Puig, Daniela Leyton and Grünwald, David and Stallinga, Sjoerd and Rieger, Bernd},
	date = {2013-06},
	pmid = {23624665},
	pmcid = {PMC4149789},
}

@article{turgut_maximum_2009,
	title = {Maximum Systolic Velocity of Inferior Thyroid Artery and Thyroideal Color Doppler Flow Pattern in Hypothyroid Subjects Before and After Treatment},
	volume = {17},
	issn = {0929-6441},
	url = {https://www.sciencedirect.com/science/article/pii/S092964410960014X},
	doi = {10.1016/S0929-6441(09)60014-X},
	pages = {44--51},
	number = {1},
	journaltitle = {Journal of Medical Ultrasound},
	shortjournal = {Journal of Medical Ultrasound},
	author = {Turgut, Ahmet Tuncay and Çakal, Erman and Koşar, Ugur and Koşar, Pınar and Demirbaş, Berrin and Aral, Yalçin},
	urldate = {2023-07-31},
	date = {2009-01-01},
	langid = {english},
}

@article{chu_morphology_2011,
	title = {The morphology and haemodynamics of the rabbit renal artery: evaluation by conventional and contrast-enhanced ultrasonography},
	volume = {45},
	issn = {1758-1117},
	doi = {10.1258/la.2011.011022},
	shorttitle = {The morphology and haemodynamics of the rabbit renal artery},
	pages = {204--208},
	number = {3},
	journaltitle = {Laboratory Animals},
	shortjournal = {Lab Anim},
	author = {Chu, Yinzhu and Liu, Haixia and Xing, Ping and Lou, Ge and Wu, Changjun},
	date = {2011-07},
	pmid = {21669903},
}

@article{errico_ultrafast_2015,
	title = {Ultrafast ultrasound localization microscopy for deep super-resolution vascular imaging},
	volume = {527},
	rights = {2015 Springer Nature Limited},
	issn = {1476-4687},
	url = {https://www.nature.com/articles/nature16066},
	doi = {10.1038/nature16066},
	pages = {499--502},
	number = {7579},
	journaltitle = {Nature},
	author = {Errico, Claudia and Pierre, Juliette and Pezet, Sophie and Desailly, Yann and Lenkei, Zsolt and Couture, Olivier and Tanter, Mickael},
	urldate = {2023-07-31},
	date = {2015-11},
	langid = {english},
}

@article{demene_spatiotemporal_2015,
	title = {Spatiotemporal Clutter Filtering of Ultrafast Ultrasound Data Highly Increases Doppler and {fUltrasound} Sensitivity},
	volume = {34},
	issn = {1558-254X},
	doi = {10.1109/TMI.2015.2428634},
	pages = {2271--2285},
	number = {11},
	journaltitle = {{IEEE} Transactions on Medical Imaging},
	author = {Demené, Charlie and Deffieux, Thomas and Pernot, Mathieu and Osmanski, Bruno-Félix and Biran, Valérie and Gennisson, Jean-Luc and Sieu, Lim-Anna and Bergel, Antoine and Franqui, Stéphanie and Correas, Jean-Michel and Cohen, Ivan and Baud, Olivier and Tanter, Mickael},
	date = {2015-11},
}

@article{jensen_anatomic_2022,
	title = {Anatomic and Functional Imaging Using Row-Column Arrays},
	volume = {69},
	issn = {1525-8955},
	doi = {10.1109/TUFFC.2022.3191391},
	pages = {2722--2738},
	number = {10},
	journaltitle = {{IEEE} transactions on ultrasonics, ferroelectrics, and frequency control},
	shortjournal = {{IEEE} Trans Ultrason Ferroelectr Freq Control},
	author = {Jensen, Jorgen Arendt and Schou, Mikkel and Jorgensen, Lasse Thurmann and Tomov, Borislav G. and Stuart, Matthias Bo and Traberg, Marie Sand and Taghavi, Iman and Oygaard, Sigrid Huesebo and Ommen, Martin Lind and Steenberg, Kitty and Thomsen, Erik Vilain and Panduro, Nathalie Sarup and Nielsen, Michael Bachmann and Sorensen, Charlotte Mehlin},
	date = {2022-10},
	pmid = {35839193},
}

@article{kang_high_2018,
	title = {High {PRF} ultrafast sliding compound doppler imaging: fully qualitative and quantitative analysis of blood flow},
	volume = {63},
	issn = {1361-6560},
	doi = {10.1088/1361-6560/aaa7a2},
	shorttitle = {High {PRF} ultrafast sliding compound doppler imaging},
	pages = {045004},
	number = {4},
	journaltitle = {Physics in Medicine and Biology},
	shortjournal = {Phys Med Biol},
	author = {Kang, Jinbum and Jang, Won Seuk and Yoo, Yangmo},
	date = {2018-02-09},
	pmid = {29334078},
}

@inproceedings{nikolov_recursive_1999,
	location = {Caesars Tahoe, {NV}, {USA}},
	title = {Recursive ultrasound imaging},
	volume = {2},
	isbn = {978-0-7803-5722-8},
	url = {http://ieeexplore.ieee.org/document/849306/},
	doi = {10.1109/ULTSYM.1999.849306},
	eventtitle = {1999 {IEEE} Ultrasonics Symposium. Proceedings. International Symposium},
	pages = {1621--1625},
	booktitle = {1999 {IEEE} Ultrasonics Symposium. Proceedings. International Symposium (Cat. No.99CH37027)},
	publisher = {{IEEE}},
	author = {Nikolov, S. and Gammelmark, K. and Jensen, J.A.},
	urldate = {2023-07-30},
	date = {1999},
	langid = {english},
}

@misc{yan_transthoracic_2023,
	title = {Transthoracic super-resolution ultrasound localisation microscopy of myocardial vasculature in patients},
	url = {http://arxiv.org/abs/2303.14003},
	doi = {10.48550/arXiv.2303.14003},
	number = {{arXiv}:2303.14003},
	publisher = {{arXiv}},
	author = {Yan, Jipeng and Huang, Biao and Tonko, Johanna and Toulemonde, Matthieu and Hansen-Shearer, Joseph and Tan, Qingyuan and Riemer, Kai and Ntagiantas, Konstantinos and Chowdhury, Rasheda A. and Lambiase, Pier and Senior, Roxy and Tang, Meng-Xing},
	urldate = {2023-04-26},
	date = {2023-03-28},
	eprinttype = {arxiv},
	eprint = {2303.14003 [eess]},
}

@article{hingot_measuring_2021,
	title = {Measuring Image Resolution in Ultrasound Localization Microscopy},
	volume = {40},
	issn = {1558-254X},
	doi = {10.1109/TMI.2021.3097150},
	pages = {3812--3819},
	number = {12},
	journaltitle = {{IEEE} Transactions on Medical Imaging},
	author = {Hingot, V. and Chavignon, A. and Heiles, B. and Couture, O.},
	date = {2021-12},
}

@article{demeulenaere_coronary_2022,
	title = {Coronary Flow Assessment Using 3-Dimensional Ultrafast Ultrasound Localization Microscopy},
	volume = {15},
	issn = {1936-878X},
	url = {https://www.sciencedirect.com/science/article/pii/S1936878X22001401},
	doi = {10.1016/j.jcmg.2022.02.008},
	pages = {1193--1208},
	number = {7},
	journaltitle = {{JACC}: Cardiovascular Imaging},
	shortjournal = {{JACC}: Cardiovascular Imaging},
	author = {Demeulenaere, Oscar and Sandoval, Zulma and Mateo, Philippe and Dizeux, Alexandre and Villemain, Olivier and Gallet, Romain and Ghaleh, Bijan and Deffieux, Thomas and Deméné, Charlie and Tanter, Mickael and Papadacci, Clément and Pernot, Mathieu},
	urldate = {2023-04-11},
	date = {2022-07-01},
	langid = {english},
}

@article{heiles_ultrafast_2019,
	title = {Ultrafast 3D Ultrasound Localization Microscopy Using a 32 × 32 Matrix Array},
	volume = {38},
	issn = {1558-254X},
	doi = {10.1109/TMI.2018.2890358},
	pages = {2005--2015},
	number = {9},
	journaltitle = {{IEEE} transactions on medical imaging},
	shortjournal = {{IEEE} Trans Med Imaging},
	author = {Heiles, Baptiste and Correia, Mafalda and Hingot, Vincent and Pernot, Mathieu and Provost, Jean and Tanter, Mickael and Couture, Olivier},
	date = {2019-09},
	pmid = {30946662},
}

@article{foroozan_microbubble_2018,
	title = {Microbubble Localization for Three-Dimensional Superresolution Ultrasound Imaging Using Curve Fitting and Deconvolution Methods},
	volume = {65},
	issn = {0018-9294},
	url = {https://www.ncbi.nlm.nih.gov/pmc/articles/PMC6459186/},
	doi = {10.1109/TBME.2018.2813759},
	pages = {2692--2703},
	number = {12},
	journaltitle = {{IEEE} transactions on bio-medical engineering},
	shortjournal = {{IEEE} Trans Biomed Eng},
	author = {Foroozan, Foroohar and O’Reilly, Meaghan A. and Hynynen, Kullervo},
	urldate = {2023-04-11},
	date = {2018-12},
	pmid = {29993387},
	pmcid = {PMC6459186},
}

@article{yan_super-resolution_2022,
	title = {Super-Resolution Ultrasound Through Sparsity-Based Deconvolution and Multi-Feature Tracking},
	volume = {41},
	issn = {1558-254X},
	doi = {10.1109/TMI.2022.3152396},
	pages = {1938--1947},
	number = {8},
	journaltitle = {{IEEE} Transactions on Medical Imaging},
	author = {Yan, Jipeng and Zhang, Tao and Broughton-Venner, Jacob and Huang, Pintong and Tang, Meng-Xing},
	date = {2022-08},
}

@inproceedings{siepmann_imaging_2011,
	title = {Imaging tumor vascularity by tracing single microbubbles},
	doi = {10.1109/ULTSYM.2011.0476},
	eventtitle = {2011 {IEEE} International Ultrasonics Symposium},
	pages = {1906--1909},
	booktitle = {2011 {IEEE} International Ultrasonics Symposium},
	author = {Siepmann, Monica and Schmitz, Georg and Bzyl, Jessica and Palmowski, Moritz and Kiessling, Fabian},
	date = {2011-10},
	note = {{ISSN}: 1948-5727},
}

@article{kittler_minimum_1986,
	title = {Minimum error thresholding},
	volume = {19},
	issn = {0031-3203},
	url = {https://www.sciencedirect.com/science/article/pii/0031320386900300},
	doi = {10.1016/0031-3203(86)90030-0},
	pages = {41--47},
	number = {1},
	journaltitle = {Pattern Recognition},
	shortjournal = {Pattern Recognition},
	author = {Kittler, J. and Illingworth, J.},
	urldate = {2023-02-17},
	date = {1986-01-01},
	langid = {english},
}

@patent{brock-fisher_means_1996,
	title = {Means for increasing sensitivity in non-linear ultrasound imaging systems},
	url = {https://patents.google.com/patent/US5577505A/en},
	holder = {Hewlett Packard Co},
	type = {patentus},
	number = {5577505A},
	author = {Brock-Fisher, George A. and Poland, {McKee} D. and Rafter, Patrick G.},
	urldate = {2023-02-17},
	date = {1996-11-26},
}

@patent{hwang_two_1999,
	title = {Two pulse technique for ultrasonic harmonic imaging},
	url = {https://patents.google.com/patent/US5951478A/en},
	holder = {Advanced Technology Laboratories Inc},
	type = {patentus},
	number = {5951478A},
	author = {Hwang, Juin-Jet and Simpson, David Hope},
	urldate = {2023-02-17},
	date = {1999-09-14},
}

@article{hansenshearer_rc_fmas,
	title = {Ultrafast 3-D Ultrasound Imaging Using Row–Column Array-Specific Frame-Multiply-and-Sum Beamforming},
	volume = {69},
	issn = {1525-8955},
	doi = {10.1109/TUFFC.2021.3122094},
	pages = {480--488},
	number = {2},
	journaltitle = {{IEEE} Transactions on Ultrasonics, Ferroelectrics, and Frequency Control},
	author = {Hansen-Shearer, Joseph and Lerendegui, Marcelo and Toulemonde, Matthieu and Tang, Meng-Xing},
	date = {2022-02},
}

@article{sauvage_4d_2020,
	title = {4D Functional Imaging of the Rat Brain Using a Large Aperture Row-Column Array},
	volume = {39},
	issn = {1558-254X},
	doi = {10.1109/TMI.2019.2959833},
	pages = {1884--1893},
	number = {6},
	journaltitle = {{IEEE} Transactions on Medical Imaging},
	author = {Sauvage, Jack and Porée, Jonathan and Rabut, Claire and Férin, Guillaume and Flesch, Martin and Rosinski, Bogdan and Nguyen-Dinh, An and Tanter, Mickael and Pernot, Mathieu and Deffieux, Thomas},
	date = {2020-06},
}

@article{stanziola_asap_2018,
	title = {{ASAP}: Super-Contrast Vasculature Imaging Using Coherence Analysis and High Frame-Rate Contrast Enhanced Ultrasound},
	volume = {37},
	issn = {1558-254X},
	doi = {10.1109/TMI.2018.2798158},
	shorttitle = {{ASAP}},
	pages = {1847--1856},
	number = {8},
	journaltitle = {{IEEE} transactions on medical imaging},
	shortjournal = {{IEEE} Trans Med Imaging},
	author = {Stanziola, Antonio and Leow, Chee Hau and Bazigou, Eleni and Weinberg, Peter D. and Tang, Meng-Xing},
	date = {2018-08},
	pmid = {29994061},
}

@article{montaldo_coherent_2009,
	title = {Coherent plane-wave compounding for very high frame rate ultrasonography and transient elastography},
	volume = {56},
	issn = {1525-8955},
	doi = {10.1109/TUFFC.2009.1067},
	pages = {489--506},
	number = {3},
	journaltitle = {{IEEE} transactions on ultrasonics, ferroelectrics, and frequency control},
	shortjournal = {{IEEE} Trans Ultrason Ferroelectr Freq Control},
	author = {Montaldo, Gabriel and Tanter, Mickaël and Bercoff, Jérémy and Benech, Nicolas and Fink, Mathias},
	date = {2009-03},
	pmid = {19411209},
}

@patent{pappalardo_bidimensional_2008,
	title = {Bidimensional ultrasonic array for volumetric imaging},
	url = {https://patents.google.com/patent/WO2008083876A3/en},
	type = {patent},
	number = {{WO}2008083876A3},
	author = {Pappalardo, Massimo and Caliano, Giosue and Caronti, Alessandro and Savoia, Alessandro Stuart and Gatta, Philipp and Longo, Cristina and Bavaro, Vito},
	urldate = {2021-08-27},
	date = {2008-09-25},
	langid = {english},
}

@inproceedings{savoia_p2b-4_2007,
	title = {P2B-4 Crisscross 2D {cMUT} Array: Beamforming Strategy and Synthetic 3D Imaging Results},
	doi = {10.1109/ULTSYM.2007.381},
	shorttitle = {P2B-4 Crisscross 2D {cMUT} Array},
	eventtitle = {2007 {IEEE} Ultrasonics Symposium Proceedings},
	pages = {1514--1517},
	booktitle = {2007 {IEEE} Ultrasonics Symposium Proceedings},
	author = {Savoia, A. and Bavaro, V. and Caliano, G. and Caronti, A. and Carotenuto, R. and Gatta, P. and Longo, C. and Pappalardo, M.},
	date = {2007-10},
	note = {{ISSN}: 1051-0117},
}

@article{bouzari_curvilinear_2017,
	title = {Curvilinear 3-D Imaging Using Row-Column-Addressed 2-D Arrays With a Diverging Lens: Feasibility Study},
	volume = {64},
	issn = {1525-8955},
	doi = {10.1109/TUFFC.2017.2687521},
	shorttitle = {Curvilinear 3-D Imaging Using Row-Column-Addressed 2-D Arrays With a Diverging Lens},
	pages = {978--988},
	number = {6},
	journaltitle = {{IEEE} transactions on ultrasonics, ferroelectrics, and frequency control},
	shortjournal = {{IEEE} Trans Ultrason Ferroelectr Freq Control},
	author = {Bouzari, Hamed and Engholm, Mathias and Beers, Christopher and Stuart, Matthias Bo and Nikolov, Svetoslav Ivanov and Thomsen, Erik Vilain and Jensen, Jorgen Arendt},
	date = {2017-06},
	pmid = {28358682},
}

@patent{ferin_ultrasound_2019,
	title = {Ultrasound transducer},
	url = {https://patents.google.com/patent/US20190328360A1/en},
	holder = {Vermon {SA}},
	type = {patentus},
	number = {20190328360A1},
	author = {{FERIN}, Guillaume and Flesch, Martin and Dumoux, Marie-Coline and Voisin, David and Legros, Mathieu and Nguyen-Dinh, An},
	urldate = {2021-05-04},
	date = {2019-10-31},
}

@article{sampaleanu_top-orthogonal--bottom-electrode_2014,
	title = {Top-orthogonal-to-bottom-electrode ({TOBE}) {CMUT} arrays for 3-D ultrasound imaging},
	volume = {61},
	issn = {1525-8955},
	doi = {10.1109/TUFFC.2014.6722612},
	pages = {266--276},
	number = {2},
	journaltitle = {{IEEE} transactions on ultrasonics, ferroelectrics, and frequency control},
	shortjournal = {{IEEE} Trans Ultrason Ferroelectr Freq Control},
	author = {Sampaleanu, Alex and Zhang, Peiyu and Kshirsagar, Abhijeet and Moussa, Walied and Zemp, Roger J.},
	date = {2014-02},
	pmid = {24474133},
}

@article{demore_real-time_2009,
	title = {Real-time volume imaging using a crossed electrode array},
	volume = {56},
	issn = {1525-8955},
	doi = {10.1109/TUFFC.2009.1167},
	pages = {1252--1261},
	number = {6},
	journaltitle = {{IEEE} Transactions on Ultrasonics, Ferroelectrics, and Frequency Control},
	author = {Demore, C. E. M. and Joyce, A. W. and Wall, K. and Lockwood, G. R.},
	date = {2009-06},
}

@article{christiansen_3-d_2015,
	title = {3-D imaging using row–column-addressed arrays with integrated apodization— part ii: transducer fabrication and experimental results},
	volume = {62},
	issn = {1525-8955},
	doi = {10.1109/TUFFC.2014.006819},
	shorttitle = {3-D imaging using row–column-addressed arrays with integrated apodization— part ii},
	pages = {959--971},
	number = {5},
	journaltitle = {{IEEE} Transactions on Ultrasonics, Ferroelectrics, and Frequency Control},
	author = {Christiansen, T. L. and Rasmussen, M. F. and Bagge, J. P. and Moesner, L. N. and Jensen, J. A. and Thomsen, E. V.},
	date = {2015-05},
}

@article{seo_5a-5_2006,
	title = {5A-5 64 × 64 2-D Array Transducer with Row-Column Addressing},
	doi = {10.1109/ULTSYM.2006.32},
	journaltitle = {2006 {IEEE} Ultrasonics Symposium},
	author = {Seo, C. and Yen, J.},
	date = {2006},
}

@article{flesch_4d_2017,
	title = {4D in vivo ultrafast ultrasound imaging using a row-column addressed matrix and coherently-compounded orthogonal plane waves},
	volume = {62},
	issn = {0031-9155},
	url = {https://doi.org/10.1088%2F1361-6560%2Faa63d9},
	doi = {10.1088/1361-6560/aa63d9},
	pages = {4571--4588},
	number = {11},
	journaltitle = {Physics in Medicine and Biology},
	shortjournal = {Phys. Med. Biol.},
	author = {Flesch, M. and Pernot, M. and Provost, J. and Ferin, G. and Nguyen-Dinh, A. and Tanter, M. and Deffieux, T.},
	urldate = {2020-01-17},
	date = {2017-05},
	langid = {english},
}

@article{christensen-jeffries_3-d_2017,
	title = {3-D In Vitro Acoustic Super-Resolution and Super-Resolved Velocity Mapping Using Microbubbles},
	volume = {64},
	issn = {1525-8955},
	doi = {10.1109/TUFFC.2017.2731664},
	pages = {1478--1486},
	number = {10},
	journaltitle = {{IEEE} transactions on ultrasonics, ferroelectrics, and frequency control},
	shortjournal = {{IEEE} Trans Ultrason Ferroelectr Freq Control},
	author = {Christensen-Jeffries, Kirsten and Brown, Jemma and Aljabar, Paul and Tang, Mengxing and Dunsby, Christopher and Eckersley, Robert J.},
	date = {2017},
	pmid = {28767367},
}

@article{christensen-jeffries_vivo_2015,
	title = {In Vivo Acoustic Super-Resolution and Super-Resolved Velocity Mapping Using Microbubbles},
	volume = {34},
	issn = {1558-254X},
	doi = {10.1109/TMI.2014.2359650},
	pages = {433--440},
	number = {2},
	journaltitle = {{IEEE} Transactions on Medical Imaging},
	author = {Christensen-Jeffries, Kirsten and Browning, Richard J. and Tang, Meng-Xing and Dunsby, Christopher and Eckersley, Robert J.},
	date = {2015-02},
}

@article{leow_flow_2015,
	title = {Flow Velocity Mapping Using Contrast Enhanced High-Frame-Rate Plane Wave Ultrasound and Image Tracking: Methods and Initial in Vitro and in Vivo Evaluation},
	volume = {41},
	issn = {1879-291X},
	doi = {10.1016/j.ultrasmedbio.2015.06.012},
	shorttitle = {Flow Velocity Mapping Using Contrast Enhanced High-Frame-Rate Plane Wave Ultrasound and Image Tracking},
	pages = {2913--2925},
	number = {11},
	journaltitle = {Ultrasound in Medicine \& Biology},
	shortjournal = {Ultrasound Med Biol},
	author = {Leow, Chee Hau and Bazigou, Eleni and Eckersley, Robert J. and Yu, Alfred C. H. and Weinberg, Peter D. and Tang, Meng-Xing},
	date = {2015-11},
	pmid = {26275971},
}

@article{christensen-jeffries_super-resolution_2020,
	title = {Super-resolution Ultrasound Imaging},
	volume = {46},
	issn = {0301-5629},
	url = {http://www.sciencedirect.com/science/article/pii/S0301562919315959},
	doi = {10.1016/j.ultrasmedbio.2019.11.013},
	pages = {865--891},
	number = {4},
	journaltitle = {Ultrasound in Medicine \& Biology},
	shortjournal = {Ultrasound in Medicine \& Biology},
	author = {Christensen-Jeffries, Kirsten and Couture, Olivier and Dayton, Paul A. and Eldar, Yonina C. and Hynynen, Kullervo and Kiessling, Fabian and O'Reilly, Meaghan and Pinton, Gianmarco F. and Schmitz, Georg and Tang, Meng-Xing and Tanter, Mickael and van Sloun, Ruud J. G.},
	urldate = {2020-06-04},
	date = {2020-04-01},
	langid = {english},
}

@article{rasmussen_3-d_2015,
	title = {3-D imaging using row-column-addressed arrays with integrated apodization - part i: apodization design and line element beamforming},
	volume = {62},
	issn = {1525-8955},
	doi = {10.1109/TUFFC.2014.006531},
	shorttitle = {3-D imaging using row-column-addressed arrays with integrated apodization - part i},
	pages = {947--958},
	number = {5},
	journaltitle = {{IEEE} Transactions on Ultrasonics, Ferroelectrics, and Frequency Control},
	author = {Rasmussen, Morten Fischer and Christiansen, Thomas Lehrmann and Thomsen, Erik Vilain and Jensen, Jørgen Arendt},
	date = {2015-05},
}

@article{li_preliminary_2015,
	title = {Preliminary work of real-time ultrasound imaging system for 2-D array transducer},
	volume = {26 Suppl 1},
	issn = {1878-3619},
	doi = {10.3233/BME-151457},
	pages = {S1579--1585},
	journaltitle = {Bio-Medical Materials and Engineering},
	shortjournal = {Biomed Mater Eng},
	author = {Li, Xu and Yang, Jiali and Ding, Mingyue and Yuchi, Ming},
	date = {2015},
	pmid = {26405923},
}

@article{holbek_3-d_2016,
	title = {3-D Vector Flow Estimation With Row-Column-Addressed Arrays},
	volume = {63},
	issn = {1525-8955},
	doi = {10.1109/TUFFC.2016.2582536},
	pages = {1799--1814},
	number = {11},
	journaltitle = {{IEEE} transactions on ultrasonics, ferroelectrics, and frequency control},
	shortjournal = {{IEEE} Trans Ultrason Ferroelectr Freq Control},
	author = {Holbek, Simon and Christiansen, Thomas Lehrmann and Stuart, Matthias Bo and Beers, Christopher and Thomsen, Erik Vilain and Jensen, Jorgen Arendt},
	date = {2016},
	pmid = {27824562},
}

@article{jensen_three-dimensional_2019,
	title = {Three-Dimensional Super Resolution Imaging using a Row-Column Array},
	issn = {1525-8955},
	doi = {10.1109/TUFFC.2019.2948563},
	journaltitle = {{IEEE} transactions on ultrasonics, ferroelectrics, and frequency control},
	shortjournal = {{IEEE} Trans Ultrason Ferroelectr Freq Control},
	author = {Jensen, Jorgen Arendt and Ommen, Martin Lind and Oygard, Sigrid Husebo and Schou, Mikkel and Sams, Thomas and Stuart, Matthias Bo and Beers, Christopher and Thomsen, Erik Vilain and Larsen, Niels Bent and Tomov, Borislav Gueorguiev},
	date = {2019-10-21},
	pmid = {31634831},
}

@article{chen_column-row-parallel_2018,
	title = {A Column-Row-Parallel Ultrasound Imaging Architecture for 3-D Plane-Wave Imaging and Tx Second-Order Harmonic Distortion Reduction},
	volume = {65},
	issn = {0885-3010, 1525-8955},
	doi = {10.1109/TUFFC.2018.2811393},
	pages = {828--843},
	number = {5},
	journaltitle = {{IEEE} Transactions on Ultrasonics, Ferroelectrics, and Frequency Control},
	author = {Chen, Kailiang and Lee, Byung Chul and Thomenius, Kai E. and Khuri-Yakub, Butrus T. and Lee, Hae-Seung and Sodini, Charles G.},
	date = {2018-05},
}
\end{document}